\documentclass[fleqn,usenatbib]{mnras}

\usepackage{newtxtext,newtxmath}
\usepackage[T1]{fontenc}
\DeclareRobustCommand{\VAN}[3]{#2}
\let\VANthebibliography\thebibliography
\def\thebibliography{\DeclareRobustCommand{\VAN}[3]{##3}\VANthebibliography}
\usepackage{graphicx}	
\usepackage{amsmath}	
\usepackage[version=4]{mhchem}

\newcommand{\tgr}{\ensuremath{\tau_{\mathrm{grey}}}}
\newcommand{\artis}{\textsc{Artis} }
\newcommand{\wl}[2]{${#1}\mathord{,}{#2}\,\text{\AA}$}

\title[Light Curves and Spectra for a 3D SN Model]{Synthetic Light Curves and Spectra for
the Photospheric Phase of a 3D Stripped-Envelope Supernova Explosion Model}

\author[T.~Maunder et al.]{Thomas Maunder$^{1}$\thanks{E-mail: 
thomas.maunder@monash.edu},
Fionntan P. Callan$^{2}$,
Stuart A. Sim$^{2}$,
Alexander Heger$^{1}$,
Bernhard M\"uller$^{1}$\thanks{E-mail: bernhard.mueller@monash.edu}
\\
$^{1}$
School of Physics and Astronomy, 10 College Walk, Monash University, Clayton, VIC 3800, Australia\\
$^{2}$School of Mathematics and Physics, Queen's University Belfast, University Road, Belfast BT7 1NN, UK \\
}
\date{Accepted XXX. Received YYY; in original form ZZZ}
\pubyear{2024}

\begin{document}
 \label{firstpage}
\pagerange{\pageref{firstpage}--\pageref{lastpage}}
\maketitle

\begin{abstract}
We present synthetic light curves and spectra from three-dimensional (3D) Monte Carlo radiative transfer simulations based on a 3D core-collapse supernova explosion model of an ultra-stripped $3.5\,\mathrm{M}_{\odot}$ progenitor. Our calculations predict a fast and faint transient with $\Delta m_{15} \sim 1\texttt{-} 2\,\mathrm{mag}$ and peak bolometric luminosity between $-15.3\,\mathrm{mag}$ and $-16.4\,\mathrm{mag}$. Due to a large-scale unipolar asymmetry in the distribution of  $^{56}\mathrm{Ni}$, there is a pronounced viewing-angle dependence with about  $1\,\mathrm{mag}$ difference between the directions of highest and lowest luminosity. The predicted spectra for this rare class of explosions do not yet match any observed counterpart. They are dominated by prominent Mg~II lines, but features from O, C, Si, and Ca are also found. In particular, the O~I line at \wl{7}{774} appears as a blended feature together with Mg~II emission. Our model is not only faster and fainter than the observed Ib/c supernova population, but also shows a correlation between higher peak luminosity and larger $\Delta m_{15}$ that is not present in observational samples. A possible explanation is that the unusually small ejecta mass of our model accentuates the viewing-angle dependence of the photometry. We suggest that the viewing-angle dependence of the photometry may be used to constrain asymmetries in explosion models of more typical stripped-envelope supernova progenitors in future.
\end{abstract}
\begin{keywords}
supernovae: general -- radiative transfer -- hydrodynamics
\end{keywords}

\section{Introduction}
Most massive stars 
are destined to end their life in a core-collapse supernova explosion after the exhaustion of the nuclear fuel in their core. Thanks to the advent of large-scale transient surveys, thousands of these events are now detected every year.
Samples of well-observed supernovae \citep[e.g.,][]{li_11,Drout2011,mueller_t_17,Taddia2018,Prentice2019} have already yielded significant insights on the systematics of supernova populations.  For some nearby core-collapse supernovae, progenitors have been directly identified using archival images \citep{van_dyk_03,smartt_09,smartt_15}. 
Observations of supernova light curves and spectroscopy cannot only probe the ``explodability''
of progenitors and bulk explosion parameters like
the explosion energy or the massive of radioactive
$^{56}\mathrm{Ni}$ made in the explosions, but have also yielded many hints about the ejecta structure and
dynamics, showing core-collapse supernovae to be inherently aspherical. Examples include evidence
for mixing processes from the observations of fast
iron clumps in SN~1987A \citep{HLi1993, chugai_88, Fryxell1991a, erickson_88}, and polarisation
signatures that constrain the geometry and degree
of asymmetries in the ejecta
\citep{WangWheeler2008, Tanaka2012, Tanaka2017}.
In recent years, multi-wavelength observations of
supernova remnants such as Cas~A and W49B have also yielded significant insights on the ejecta structure and composition \citep{DeLaney2010, Isensee2010, Lopex2013, grefenstette2014}.

At the same time, the last few years have seen
tremendous progress in our theoretical understanding of the core-collapse supernova explosion mechanism now
that multi-dimensional fluid flow, long recognised to be critical in the explosion mechanism \citep{Muller2020}, can be adequately modelled in models with sophisticated
neutrino transport. Many three-dimensional (3D) simulations have been able to obtain explosions for a wide range of supernova progenitors \citep[e.g.,][]{takiwaki_12, melson_15a, melson_15b, mueller_15b, mueller_19a, vartanyan_19, burrows_19, Powell2024}. The successes
are not limited to merely modelling shock revival,
simulations with detailed physics have become sufficiently mature to allow for the calculation of explosion energy, nickel masses, compact remnant masses, kicks, and spins \citep{mueller_17,mueller_19a,nakamura_19, bollig_21,burrows_24} as key explosion and remnant properties that can be compared to observationally inferred values.

The most ambitious approach to test and validate the modern picture of core-collapse supernova explosions relies on \emph{synthetic observables}, ideally from
first-principle simulations that can be pitched directly against observational data. This has
long been challenging, as this required extending
3D models from the short engine phase (which lasts for only several seconds) beyond shock breakout on much longer time scales.
 During shock propagation through the stellar envelope, multi-dimensional effects also play a
 major role as  
 Rayleigh-Taylor \citep{Chevalier1976}, Kelvin-Helmholtz, and Richtmyer-Meshkov \citep{richtmyer1960} instabilities
 reshape the spatial distribution of the ejecta compared to the initial asymmetries seeded during the engine phase \citep{WangWheeler2008, Muller2020}.

Simulations of mixing instabilities in supernova envelopes have advanced significantly (see \citealt{Muller2020} for a review) since the seminal 2D simulations of instabilities and clumping in the aftermath of SN~1987A \citep{arnett1989, mueller1991, Fryxell1991a, Hachisu1991, Benz1990}. The most sophisticated modern 3D simulations of mixing instabilities start from 3D engine models with parametrised neutrino transport and tuned engines \citep{Hammer2010, wongwathanarat_13, Wongwathanarat2015, Wongwathanarat2017}, or even from explosion models with multi-group neutrino transport \citep{chan_18, chan_20}. So far many of these studies have focused on nearby events or their remnants such as SN~1987A \citep{Hammer2010, wongwathanarat_13, Wongwathanarat2015, Wongwathanarat2017, Larsson2013} and Cas~A \citep{Wongwathanarat2017}.

In choosing further points of reference for a comparison between
models and observations and constraining the explosion mechanism, it is important to consider the
disparity between engine asymmetries and the observable asymmetries, which limits conclusions on the multi-dimensional
dynamics of the engine. For this reason, \emph{stripped-envelope} supernovae are of particular interest for studying explosion asymmetries. In red and blue supergiant progenitors, strong mixing by the Rayleigh-Taylor instability is caused by the acceleration and subsequent deceleration of the shock at the interface between the helium and hydrogen shell.
In stripped-envelope supernovae, the smaller extent or complete lack of the hydrogen envelope implies that initial phase asymmetries are not reshaped as much by strong Rayleigh-Taylor mixing as in explosions of red or blue supergiants.

Type Ib/c supernovae from hydrogen-free progenitors are therefore particularly promising for leveraging explosion asymmetries as diagnostics of the supernova engine. Observational diagnostics for explosion asymmetries in stripped-envelope supernovae
have already been studied extensively. Evidence for non-axisymmetric ejecta in normal Type Ib/c supernovae (as opposed to broad-line Type Ic supernovae), possibly indicative of convective plumes as seed asymmetries in neutrino-driven mechanism, has been found, e.g. by \citet{Tanaka2012, Tanaka2017} in the form of Q-U loops in the Stokes diagram.  The colour evolution in Type Ib/c supernovae has been used as a probe for mixing by correlating the early-time colour evolution of Type Ib/c's with \ce{^56Ni} mixing \citep{Yoon2019}. 
Line profiles from
nebular emission also have a long history as diagnostics for
asphericity in Type Ib/c ejecta \citep{Taubenberger2009,Fang2022, Milisavljevic2010}.

The requisite simulations that can be compared to the rich observational evidence on stripped-envelope supernovae are now becoming available.
Early 2D simulations \citep{Hachisu1991, Hachisu1994, kifonidis_00, Kifonidis2003} used progenitors with their hydrogen envelopes artificially removed instead of consistently computing the stellar structure and evolution of Type Ib/c supernovae progenitors. 
Modern multi-D simulations with grey \citep{van_baal_23} or multi-group neutrino transport \citep{Muller2018,mueller_19a} can be performed using more consistent progenitor models based on binary stellar evolution, which is crucial for properly capturing the mixing dynamics in the envelope.
Recently, such simulations have been followed
up until and beyond shock breakout and used
to compute synthetic observables using Monte
Caro radiative transfer for the first time.
\citet{van_baal_23} presented a parameterised long-time 3D simulations of  a stripped-envelope Type Ib supernova with pre-determined explosion energies and then used this for 3D radiative transfer calculations during the \textit{nebular} phase.
Synthetic light curves and spectra for the \emph{photospheric phase} based on a self-consistent two-dimensional (2D) explosion model and multi-D Monte Carlo radiative transfer (MCRT) were first presented in our previous work on ultra-stripped supernovae \citep{Maunder2024}.

In this paper, we extend the work of \citep{Maunder2024} on the photospheric phase to 3D and consider a less extreme progenitor model with higher explosion energy and ejecta mass.
The study of
\citet{Maunder2024} considered a very extreme case
with with a uniquely thin helium envelope and a
a core structure that facilitated a rapid explosion, all of which simplified simulations to shock breakout. Moreover, the progenitor model used by \citet{Maunder2024} underwent a pre-collapse mass ejection, so that interaction significantly affects
the light curves \citep{moriya2025}, making this model rather non-representative of typical Ibc supernovae. Instead we use a model with a helium envelope of $0.52\,\mathrm{M}_\odot$. While the model still lies a the very low-mass end of the spectrum of ejecta masses in Ibc supernovae, it is somewhat more suitable to a comparison with observed events.

We use three-dimensional Monte Carlo radiative transfer to calculate synthetic light curves and spectra based on a first-principle explosion model of a stripped-envelope supernova from \citet{mueller_19a}. This allows us to investigate the role of ejecta asymmetries on observable light curves and spectra. 
In particular, we consider the viewing-angle dependence for our 3D model, which shows significant variations in peak luminosity and light curve shape for different observer directions.
We compare the model with the population of observed Type Ib/c supernovae to determine the viability of the model and diagnose tensions of current models with observed Type Ib/c supernovae.

\section{Progenitor and Explosion Model}
In this work, we compute synthetic light curves and spectra for the explosion of a $3.5\,\mathrm{M}_{\odot}$ ``ultra-stripped'' helium star \citep{Tauris2015} with an initial metallicity of $\mathrm{Z}=0.02$. The progenitor evolution was simulated using the binary evolution code \textsc{bec} until oxygen burning. In order to reach collapse, the progenitor was then mapped from the binary evolution code \textsc{bec} to the stellar evolution code \textsc{Kepler} \citep{Weaver1978, Heger1999}.

\begin{figure}
    \centering
    \includegraphics[width=\linewidth]{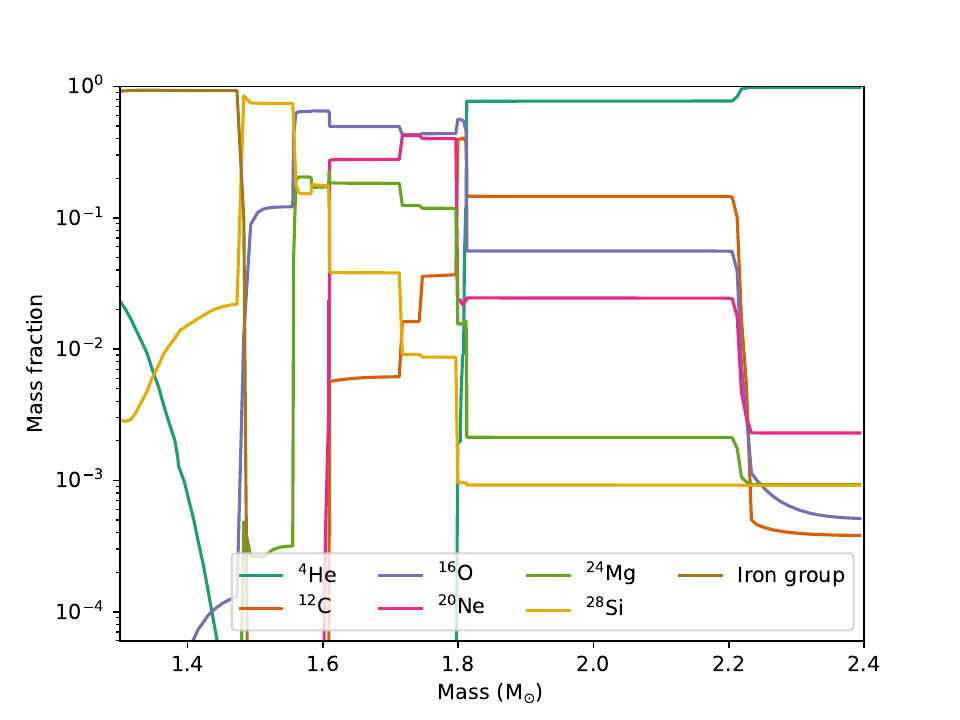}
    \caption{Pre-supernova composition of the progenitor before the onset of core collapse. Only selected elements in the outer part of the iron core and the outer shell are shown.}
    \label{fig:progen_comp}
\end{figure}

The star underwent a Case BB common envelope event  \citep{Wellstein2001, Yoon2010}. After the common envelope event, the progenitor was left with a mass of $2.39\,\mathrm{M}_{\odot}$, which is about
the mass that remains at the onset of collapse as well.  The composition of the
progenitor at the onset of collapse is shown
in Figure~\ref{fig:progen_comp}.
It has an Fe-Si core of almost $1.5\,\mathrm{M}_\odot$,
followed by an active oxygen burning shell, an inert
O-Ne-Mg shell, a Ne burning shell, and an almost completely
burned C shell. Further out, there is a thick He burning
shell from $1.8\,\mathrm{M}_\odot$ to $2.2\,\mathrm{M}_\odot$,
followed by a He envelope of about
$0.2\,\mathrm{M}_\odot$.

The collapse and explosion were simulated in 3D using the relativistic neutrino radiation hydrodynamics code \textsc{CoCoNuT-FMT} \citep{Muller2015}. At $1\,\mathrm{s}$ after core bounce, the model was mapped
to the \textsc{Prometheus} code \citep{Fryxell1991a,mueller1991} to 
follow the evolution beyond shock breakout
following the methodology of \citet{Muller2018}. A moving grid with an initial radial resolution of $1600$ zones
was used for the simulation beyond shock breakout. The angular resolution was  $1.6^\circ$ on an overset, axis-free Yin-Yang grid \citep{wongwathanarat_10}. 
Both \textsc{CoCoNuT-FMT} and \textsc{Prometheus}
use the consistent multi-fluid advection method of
\citet{plewa_99} to enforce strict conservation of partial
masses during advection while maintaining mass fractions
that sum up exactly to unity everywhere.

\begin{figure}
    \centering
    \includegraphics[width=\linewidth]{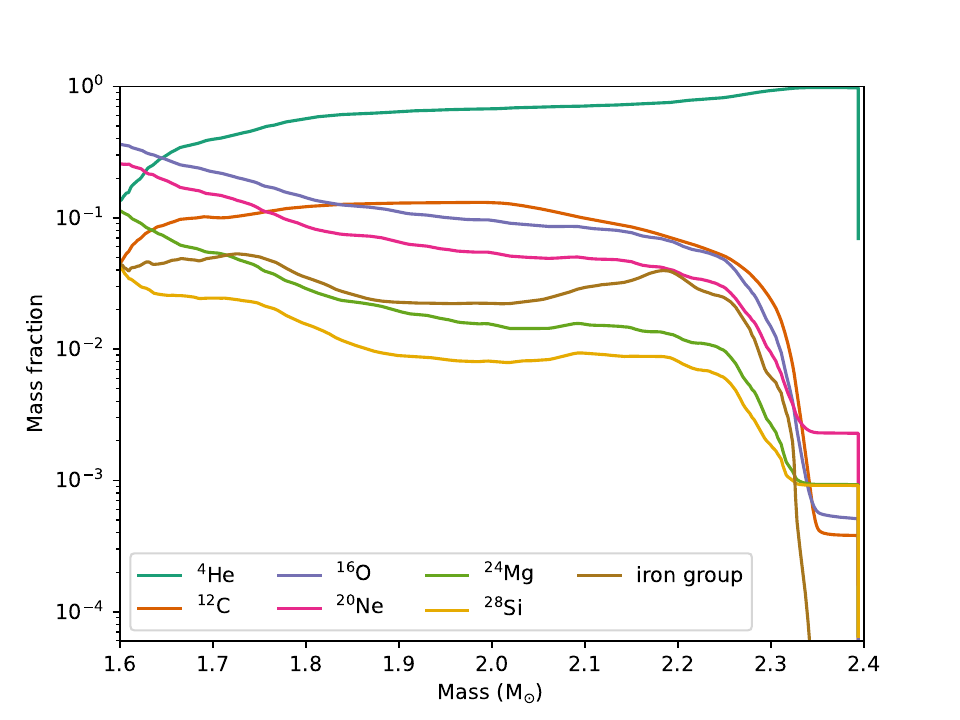}
    \caption{Spherically averaged composition of the 3D explosion model before mapping to the 3D
    Cartesian grid in \artis, 
    showing how}
    mixing instabilities have modified the initial shell structure of the progenitor.
    \label{fig:prom_comp}
\end{figure}

The model reached a diagnostic energy of $2.78\times10^{50}\,\mathrm{erg}$ by the end of the \textsc{CoCoNuT} simulation. During the propagation
of the shock to the envelope, the energy decreases slightly
to a final value$2.5\times10^{50}\,\mathrm{erg}$ as the shock sweeps up the envelope material, which still has a non-negligible binding energy (``overburden'').
About $0.8\,\mathrm{M}_\odot$ of material is ejected; the innermost $\mathord\sim 1.6\,\mathrm{M}_\odot$ end up in the neutron star \citep{mueller_19a}. About $0.023\,\mathrm{M}_\odot$ of iron-group material is ejected, of which $0.0083\,\mathrm{M}_\odot$ is
directly classified as $^{56}\,\mathrm{Ni}$ in the code.
The version of the \textsc{CoCoNuT-FMT} code tended to produce slightly too neutron-rich inner ejecta \citep{sieverding}, so much of the remaining, slightly neutron-rich iron-group material, is likely to form $^{56}\,\mathrm{Ni}$ in reality. The ejecta also contain $0.52\,\mathrm{M}_\odot$ of helium, 
$0.071\,\mathrm{M}_\odot$ of carbon,
$0.095\,\mathrm{M}_\odot$ of oxygen,
$0.021\,\mathrm{M}_\odot$ of magnesium, $0.060\,\mathrm{M}_\odot$ of neon,
$0.010\,\mathrm{M}_\odot$ of silicon.
The spherically averaged composition of the explosion model
at a time of $3424.49\,\mathrm{s}$ is shown in Figure~\ref{fig:prom_comp}.

\section{Methods}
Synthetic light curves and spectra for the 3D explosion model were generated using the Monte Carlo radiative transfer (MCRT) code \artis \citep{sim2007b, Kromer2009, Bulla, Shingles2020}.
The Monte Carlo packets provide frequency-dependent information on the radiation field, from which the emerging spectra and, by integration over frequency or frequency bands, light curves can be constructed.
For a comparison to other photon radiative transfer codes for supernova explosions, see \citet{blondin_22}.
The code assumes the ejecta are in homologous expansion and that radioactive heating from the decay of ${}^{56}\mathrm{Ni}$ and ${}^{56}\mathrm{Co}$ is the only power source for the transient. 
The decay chains of $^{52}\mathrm{Fe}$ and
$^{48}\mathrm{Cr}$ are also included in \artis, but their contribution to radioactive heating is negligible for the purpose of this study.
Additional power sources such as interactions, magnetised winds, or other central engines are not included. For the earliest phase of supernova
light curves, i.e., during and shortly after shock breakout, a true radiation-hydrodynamics treatment would be required. For the ultra-stripped envelope model discussed in \citet{Maunder2024}, \citet{moriya2025} found noticeable
differences in the light curve before peak between \artis and the
radiation hydrodynamics code \textsc{Stella} \citep{blinnikov_98} because the
low mass and large radius of the tenuous helium envelope of the progenitor led to a very bright shock breakout and an extremely rapid evolution of the light curve. The progenitor considered here does not have such an extreme envelope structure, so that \artis can be relied upon around peak and during the later part of the rise phase.

\artis features a comprehensive set of relevant radiative processes. 
Our \artis simulations include a detailed set of bound-bound transitions from elements with atomic number $Z=2$ and $Z=6$ to $30$ (drawn from the CD23 atomic data set, as described by 
\citealt{Kromer2009}, and treated using the Sobolev approximation), as well as bound-free, free-free and electron scattering opacities.

To expedite radiative transfer simulations,
\artis applies a grey approximation for the transport in very optically thick cells, whereby packets in zones at high optical depth, defined as having grey optical depth higher than some threshold value, \tgr\ are  assumed only to undergo grey scattering and no other interactions with matter (i.e. this has the effect that, at high optical depth, the radiation packets are simply advected with the flow while losing energy due to the expansion; the grey opacity coefficient used is calculated using Equation~(8) of \citealt{Sim2007}).We determined that a high value of $\tgr= 2500$ is required for convergence.

Note that this study employs the approximate non-LTE treatment of \citet{Kromer2009}, which is expected to be reasonably accurate for the early phase up to and around peak light, but does not include full non-LTE and non-thermal particle excitation/ionisation as required for late-phase spectra. For late-phase spectra, the upgraded non-LTE treatment of \citet{Shingles2020} would be required, but is not included in this study as we focus on the light curves and spectra around peak. Adopting the approximate non-LTE treatment does, however, mean that our current simulations cannot accurately determine whether spectral lines of helium will form, since modelling the excitation of states responsible for optical/NIR transitions of helium depends on non-thermal, non-LTE processes \citep[see, e.g.,][]{lucy_91}. Omitting a detailed treatment of the excitation/ionization of helium is not expected to dramatically alter the general features of our spectra or light curves

as \artis includes helium contributions to opacity and
helium ionisation that provides free electrons, but it does mean that we cannot draw strong conclusions on whether (or when) potential optical/NIR spectral lines of helium might appear.

The hydrodynamic explosion model was mapped from the 3D Yin-Yang grid to a 3D Cartesian grid (as required by \artis) with $100\times 100\times 100$ zones, which is finer than in typical applications of \artis to Type~Ia supernovae or kilonovae. For the mapping, we implemented a conservative algorithm that iteratively subdivides the cells of the Yin-Yang grid until the subcells either lie completely within a cell of the Cartesian target grid, or are smaller than a pre-specified fraction of the target cell. The subdivided cells of the original grids are then completely assigned to the target cell that contains their cell centre coordinates. This ensures exact conservation of the total mass and the masses of individual elements and nuclides. Second-order minmod reconstruction is used during the subdivision process to improve the smoothness of the mapped density and mass fraction fields while avoiding overshooting.

After mapping, homologous expansion is assumed.
The radiative transfer simulation therefore need not start at the exact time of mapping.
In this study we map the hydrodynamical
simulation into \textsc{Artis} at $3\mathord,424\,\mathrm{s}$ after the onset of the explosion, assume homologous adiabatic expansion until 1 day, and then run the radiative
transfer simulations from 1 to 120 days.

Similar to our previous work \citep{Maunder2024},
the treatment of the ejecta composition in the radiative transfer simulation bears careful consideration. The explosion model only tracks 20 nuclear species, namely, \ce{^1H}, \ce{^3He}, \ce{^12C}, \ce{^14N}, \ce{^16O}, \ce{^20Ne}, \ce{^24Mg}, \ce{^28Si}, \ce{^32S}, \ce{^36Ar}, \ce{^40Ca}, \ce{^44Ti}, \ce{^48Cr}, \ce{^52Fe}, \ce{^54Fe}, \ce{^56Ni}, \ce{^56Fe}, \ce{^60Fe}, \ce{^62Ni}, along with protons and neutrons. Nuclear burning is only treated with a simple ``flashing'' treatment based on threshold temperatures \citep{Rampp2000}, and nuclear statistical equilibrium (NSE) is assumed above
a threshold temperature of $5\, \mathrm{GK}$.
Material undergoing freeze-out from NSE will retain its composition at the threshold temperature. This burning treatment is adequate
for approximately capturing the overall yields from key burning regimes within reasonable accuracy (e.g., the overall amount of iron group material from Si burning, etc.), but cannot predict the detailed composition of the iron group ejecta self-consistently. The simple burning treatment also ignores potentially important non-$\alpha$ elements (N, Na, etc.) below the iron group. The iron group composition is particularly relevant for the powering of the light curve via the decay of $^{56}\mathrm{Ni}$ and $^{56}\mathrm{Co}$, but since the isotopic composition is sensitive to the electron fraction in the inner ejecta and hence to details in the neutrino transport, appropriate parameterisations of the composition of the iron-group ejecta can be justified.

Abundances missing from the explosion model are therefore set by modifying the mapped composition based either on typical yields from explosive burning, or on the progenitor composition in unburned regions. 
The procedure works as follows.
When mapping from \textsc{Prometheus} to \artis and protons, neutrons are all lumped into 
\ce{^1H}.
For the species \ce{^2He} through \ce{^36Ar}, we directly use the abundances from the explosion model. Abundances of lithium, beryllium, and boron are set to zero. The mass fractions of
species from atomic number 20
or above (starting form \ce{^40Ca}) are determined by re-scaling their solar abundances to fit the total iron-group mass fraction in the \textsc{Prometheus} simulation in any given grid cell. This procedure ensures that  the abundance \emph{ratios} in the synthesised iron-group material are close to their solar values after the nuclear decays. In iron-rich ejecta, the solar abundances of these species are scaled up proportionally, as one expects them to be made with similar production factors in core-collapse supernovae. The unstable isotope $^{56}\mathrm{Co}$  
and its stable daughter nucleus $^{56}\mathrm{Fe}$ are taken to originate only from the decay of \ce{^56Ni}, and their abundances are set according to the elapsed radioactive decays from the start of the explosion. For Ni, we also include  \ce{^58Ni} separately from \ce{^56Ni}  according to (scaled) solar abundances. \ce{^57Ni} is not included.
The \textsc{Prometheus} simulation includes a dummy nuclear species for material from the numerical atmosphere around the star. This material is not included when mapping 
to \textsc{Artis}. The mass fractions therefore need to
be renormalised to unity after mapping.

We here present results from two full 3D radiative transfer simulations. The radiative transfer simulation using the mapped 3D structure
and abundances (with the above modifications)
is our baseline model. In addition, we consider
a model with a increased mass of $^{56}\,\mathrm{Ni}$ as part of a sensitivity
analysis, see Section~\ref{sec:photo_avg} for details and discussion.

\section{Photometry}
\label{sec:photo}
\subsection{Angled-averaged Emission}
\label{sec:photo_avg}
\begin{figure*}
    \centering
    \includegraphics[width=\linewidth]{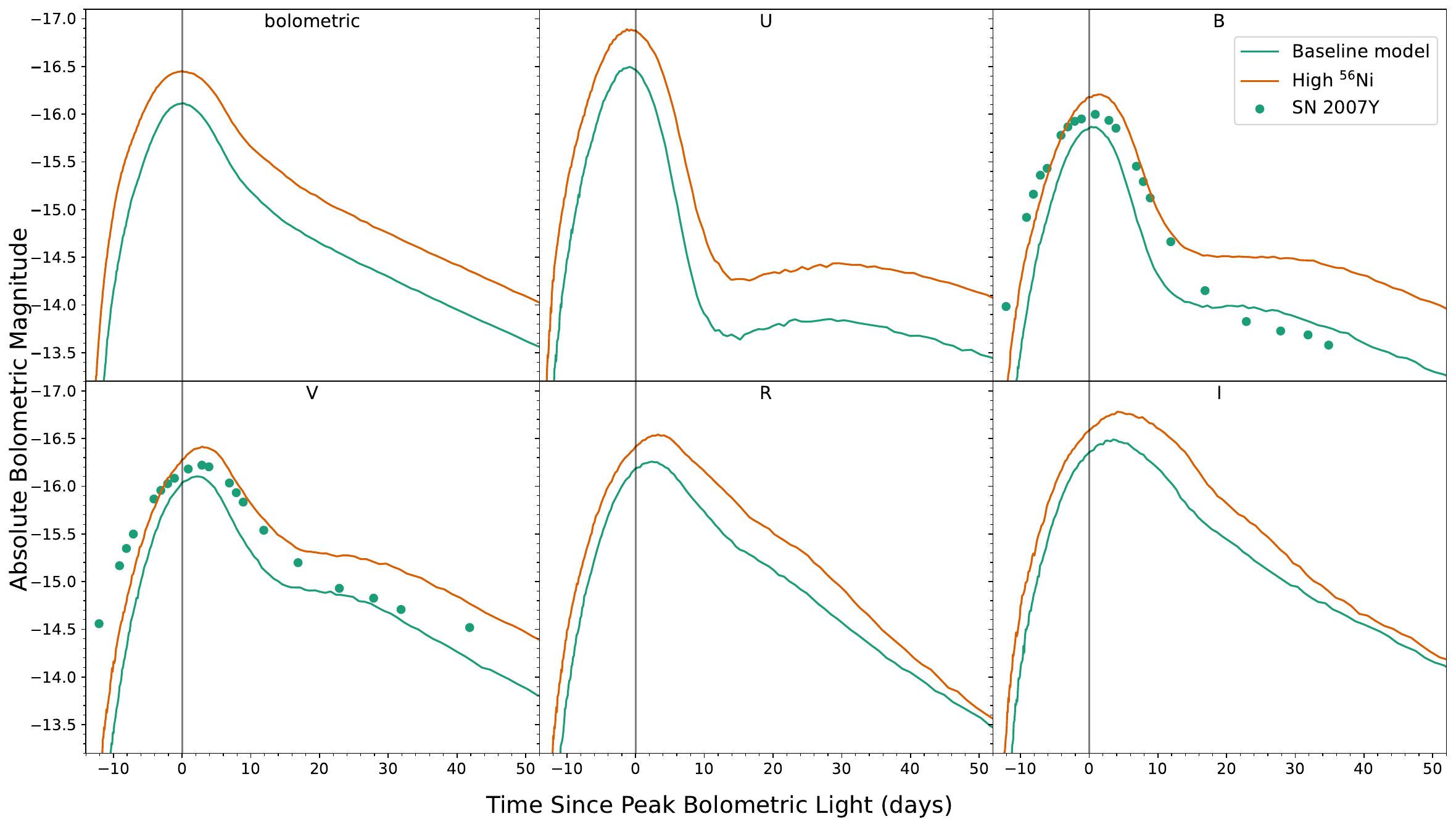}
    \caption{Comparison of the bolometric and UBVRI light curves showing the comparison to the model with increased nickel mass. 
    Orange curves show the results for the baseline model.}
    The green curve investigated the sensitivity to stronger powering by \ce{^56Ni} where we increased the mass of \ce{^56Ni} by 30\%, resulting in an increase in peak luminosity of approximately 50\%, which is in agreement with what we would expect from Arnett's rule but does not make our model bright enough to match many more observed events.
    B- and V-band photometry for
    SN~2007Y \citep{Stritzinger2009} is shown as green dots for comparison (see Section~\ref{sec:pop_comp} for discussion).
    \label{fig:cband_comp}
\end{figure*}

We show the bolometric light curves as well as light curves in the U, B, V, R, and I band in Figure~\ref{fig:cband_comp}.
These light curves are averages over viewing
angles, i.e., they are constructed from the genuinely
three-dimensional baseline model using all
Monte Carlo packets emerging in any direction.
Light curves are shown both for the baseline model and for a case with a manually increased mass of $^{56}\mathrm{Ni}$
Note that the U-, B-, and V-band show a bump or plateau after the peak before transitioning to an
exponential decline in the tail phase.
The bump is vaguely similar to a feature in some of Type~Ia merger models, see, e.g., Figure 6~
\citet{pakmor_22}, who tentatively ascribed to recombination for Fe-peak elements, which impacts the fluorescent redistribution between different bands.

The baseline model reaches a peak bolometric luminosity of $8.4\times10^{41}\,\mathrm{erg}\,\mathrm{s}^{-1}$ after 15 days, putting it distinctly in the faint end of the distribution of observed stripped-envelope supernovae. 
The UBVRI peak magnitudes (Table~\ref{tab:lc_spherical})
range from
-16.49 in U band to -15.87 in B band. These are considerably fainter than the average magnitudes in the 
non-volume limited sample of stripped-envelope
supernovae of \citet{Drout2011}, who find 
average R-band peak magnitudes of $-17.0\pm 0.7$
for Ib supernovae and $-17.4\pm 0.4$ for Ic supernovae.
The peak magnitudes are, however, quite
similar to the average value in the volume-limited
LOSS survey \citep{li_11} of about $-16.1$ for Ib/c supernovae.

\begin{table}
    \centering
    \begin{tabular}{c|c|c|c|c}
    \hline
    \hline
        Band & Peak magnitude & Peak time \\
         & (mag) & (days)\\
        \hline
        U & -16.49 & 14\\
        B & -15.87 & 15\\
        V & -16.10 & 17\\
        R & -16.25 & 17\\
        I & -16.49 & 18\\
    \hline
    \hline
    \end{tabular}
    \caption{Peak magnitudes and time of peak for different photometric bands. Data are shown for the baseline model 
    in Figure~\ref{fig:cband_comp}. }
    \label{tab:lc_spherical}
\end{table}

Multi-dimensional supernova models are still subject to uncertainties that need to be taken into account in comparisons with observed Ib/c supernovae. Specifically,
uncertainties in the explosion energy and the synthesised
mass of $^{56}\mathrm{Ni}$ will affect the light curve.
Because of the computational costs of 3D supernova simulations with multi-group neutrino transport, it is
not feasible yet to simulate multiple realisation of the explosions for a particular progenitor with different physics assumptions (e.g., closure relations in the neutrino transport, magnetic fields) that may affect these explosion outcomes.
We can, however, still explore some uncertainties by
parametrically varying the initial model for the radiative transfer simulation.
Most notably, we simulated a model with an artificially
increased nickel mass. 
Uncertainties in the nickel mass exist because of various limitations and approximations in the underlying explosion model. The freeze-out from NSE is treated very simply by switching NSE off
at a temperature of $5\,\mathrm{GK}$, when there is typically still a sizeable fraction of $\alpha$-particles around. Further recombination of $\alpha$-particles or earlier freeze-out, which could increase or decrease the nickel mass, are not taken into account and would require a nuclear reaction network. Second, uncertainties in the freeze-out composition exist because of the sensitivity of the electron fraction $Y_\mathrm{e}$ to minute details of the neutrino transport and neutrino reaction rates, although the overall amount of iron-group material is less affected by uncertainties in $Y_\mathrm{e}$. Finally, any uncertainties in the explosion energetics, e.g., due to details of the neutrino transport, the nuclear equation of state, or magnetohydrodynamic effects, translate into uncertainties in the amount of nickel produced in the neutrino-driven outflows or by explosive burning in the shock.

To increase the nickel mass while maintaining a total mass fraction of one for all cells, we first increased the nickel mass in each zone by a scale factor. After this we took the sum of each zone and used that to renormalise all species within a zone such that the mass fraction in each zone summed correctly to one. 
This procedure has the advantage of increasing the mass fraction of nickel only where it is present already, and is therefore suitable for exploring uncertainties due to NSE freeze-out conditions than also changing the spatial distribution of nickel. However, if we preserve the spatial distribution of nickel, the scaling factor is limited; we can only increase the maximum mass fraction of iron-group material\footnote{Note that the maximum mass fraction in 3D is significantly higher than after spherical averaging in Figure~\ref{fig:prom_comp}} from $\mathord{\sim}0.7\texttt{-}0.8$ to about 1. This, on the other hand roughly exhausts the range of uncertainties due to the simplified treatment of NSE freeze-out, which could at best result in pure iron-group composition without any light particles.

For computational convenience, the sensitivity test with increased
nickel mass was performed with two runs with a setting of \tgr=1000. 
The small convergence error  -- about 10\% in peak bolometric luminosity
and much less before and after peak --
is significantly smaller than the impact of the increased nickel mass.

We require 30\% more nickel mass in order to increase the peak luminosity by $0.34\,\mathrm{mag}$ or $46\,\%$\footnote{This comparison refers to the two runs with an identical setting of \tgr=1000, not to the baseline model.}.
This is a significant increase in the nickel mass for minimal luminosity increase. This is roughly in accordance with the scaling
from Arnett's rule \citep{arnett_82}.
The other bands show similar behaviour as the bolometric light curve for the sensitivity tests,
with the model with more nickel producing a brighter explosion. 
Note that the R- and I-band light curves in the model
with enhanced nickel mass converge towards the baseline
model at late times.

\subsection{Viewing-Angle Dependence}
In order to investigate potential signatures of explosion asymmetries, we next consider variations in the light curves for different viewing angles.
Although the viewing-angle dependence cannot be investigated directly for any observed supernova (except perhaps in a limited fashion in the case of light echos), such a viewing-angle dependence could be studied indirectly by considering variations in the observed supernova population,
which we shall address later in Section~\ref{sec:pop_comp}.

\begin{figure}
    \centering
    \includegraphics[width=\linewidth]{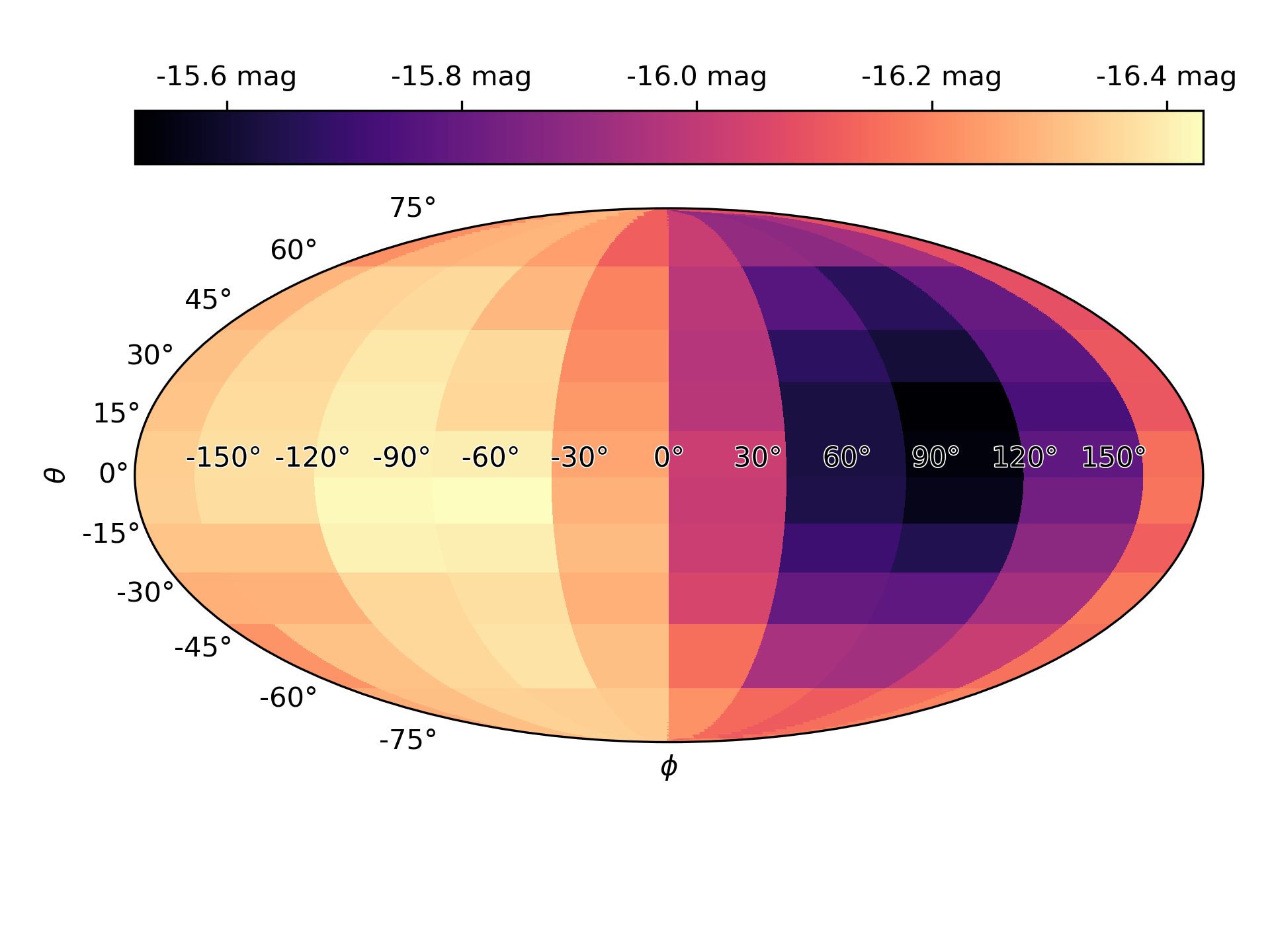}
    \caption{Peak bolometric magnitude for different observer viewing angles. This is the peak magnitude that different observers would measure at slightly different peak times
    (see Figure~\ref{fig:vangle_all_plight})
    based on the flux in their observer direction.}
    The 100 viewing angle bins
    are equally spaced in longitude $\phi$ and in the sine of latitude $\theta$, i.e., each bin covers the same solid angle.
    \label{fig:vangle_plight}
\end{figure}

\begin{figure}
    \centering
    \includegraphics[width=\linewidth]{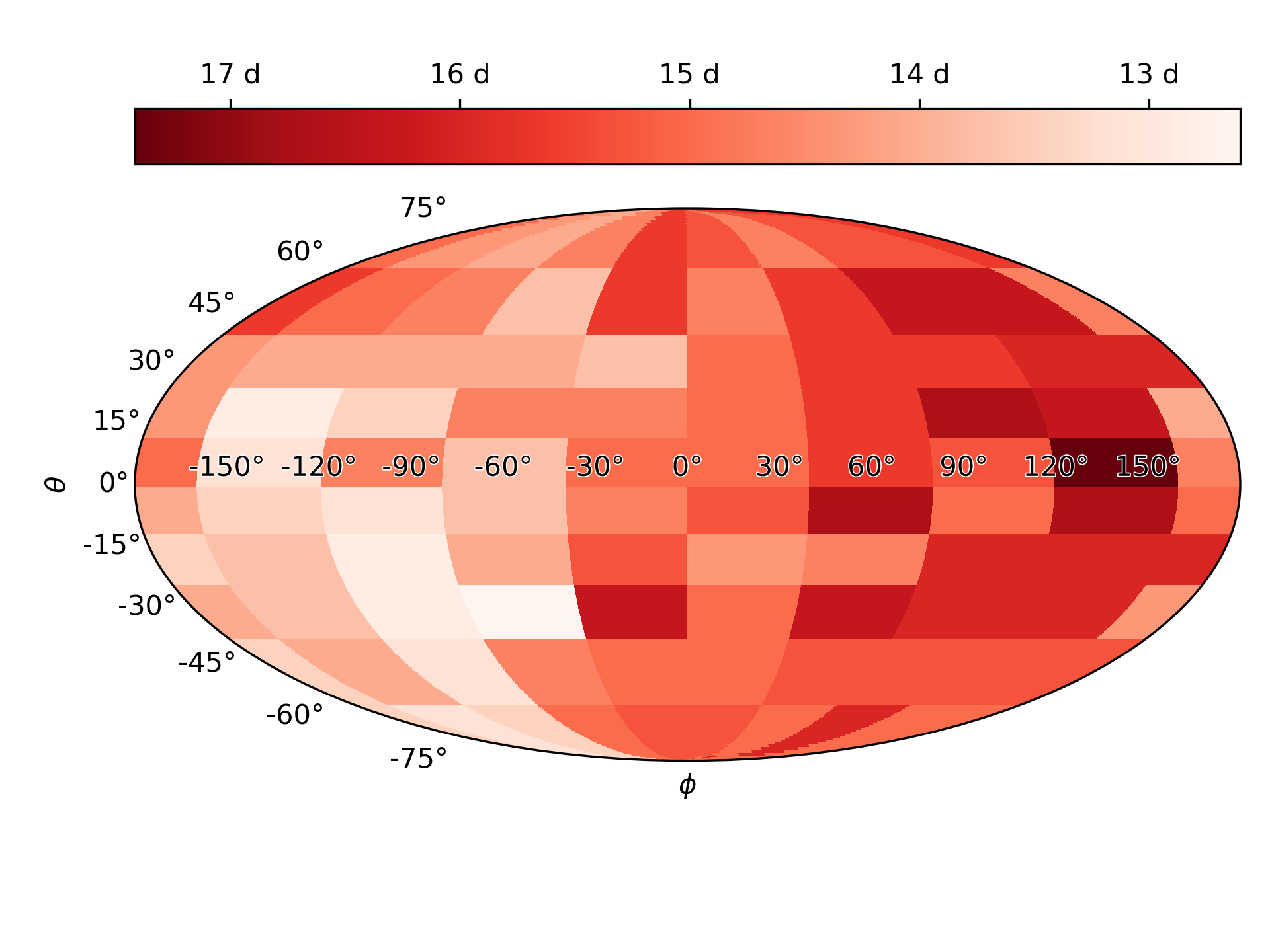}
    \caption{Time of peak bolometric luminosity as a function of viewing angle.}
    \label{fig:vangle_all_plight}
\end{figure}

We obtain viewing-angle dependent light curves by binning the emerging photon packets in the baseline radiative transfer simulations into 100 bins of equal solid angle, with uniform spacing in longitude $\phi$ and equal spacing in the cosine of colatitude $\theta$.
Figure~\ref{fig:vangle_plight} shows the viewing-angle dependence of the absolute bolometric magnitude at peak
light, which will occur at a different time for different viewing angles.
The Mollweide projections show a strong dipolar pattern in peak magnitude. The amplitude of the viewing-angle dependence is about $1\,\mathrm{mag}$.

There is also a significant viewing-angle dependence in
the time of peak light.
Figure~\ref{fig:vangle_all_plight} shows the time
of peak light in days using a similar projection as in Figure~\ref{fig:vangle_plight}. Along the viewing angles associated with brighter peak magnitudes, the peak tends to occur earlier. The time of peak varies
from 13 to 18 days with a mean of 15 days.
The temporal evolution of the brightness variation is shown in Figure~\ref{fig:vangle_all_plight}. The strong viewing-angle dependence is present during the rise and around peak luminosity, but disappears later after peak.

\begin{figure*}
    \centering
    \includegraphics[width=\linewidth]{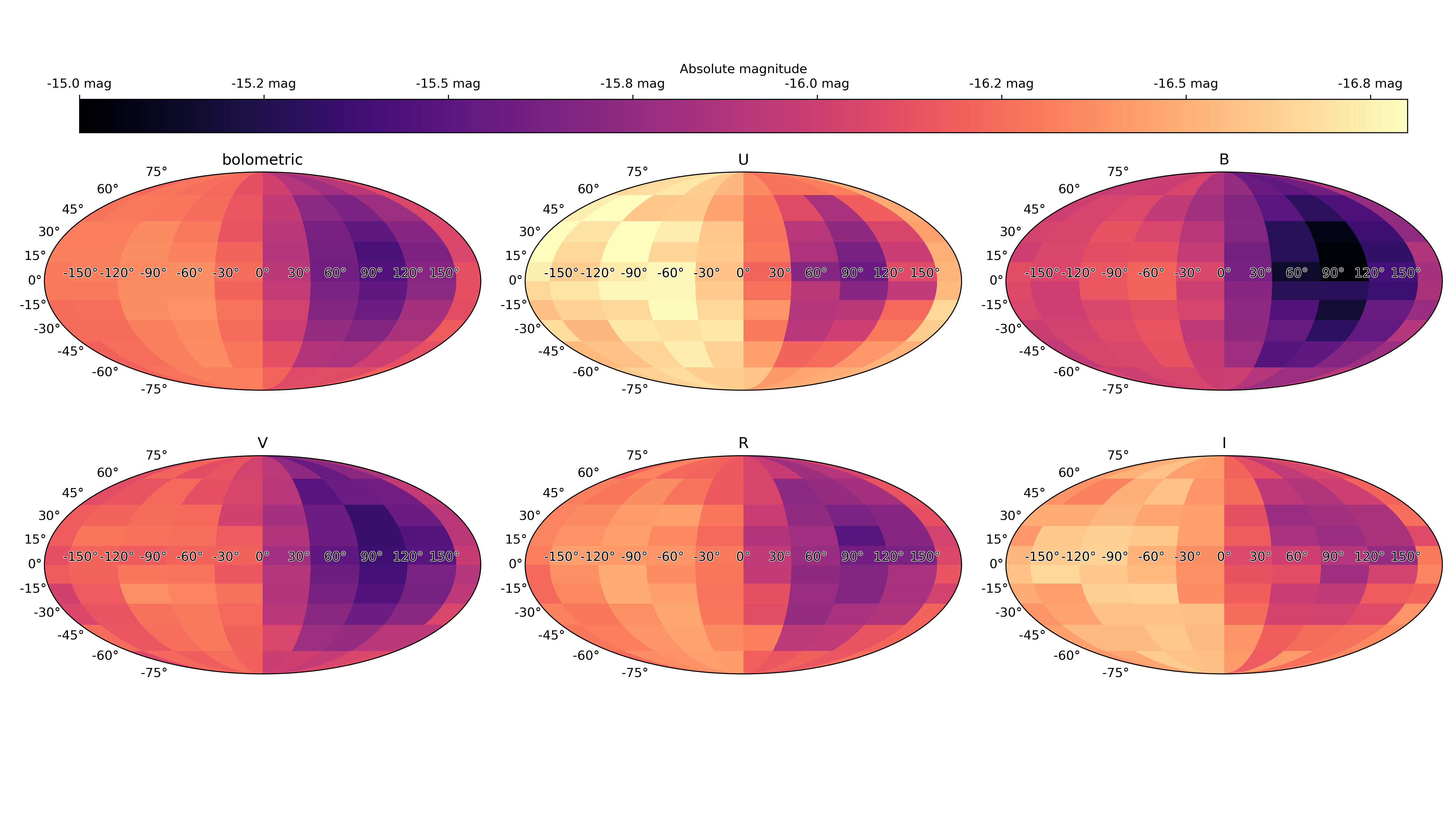}
    \caption{Viewing-angle dependence of the U, B, V, R, I bands at peak bolometric luminosity for each observer direction
    in addition to the viewing angle-dependence of bolometric magnitude. 
    All quantities are shown at the \emph{same} epoch of $14.3\,\mathrm{d}$, corresponding to the peak of the angle-averaged bolometric luminosity.}
\label{fig:multiplot_plight}
\end{figure*}

\begin{figure*}
    \centering
    \includegraphics[width=\linewidth]{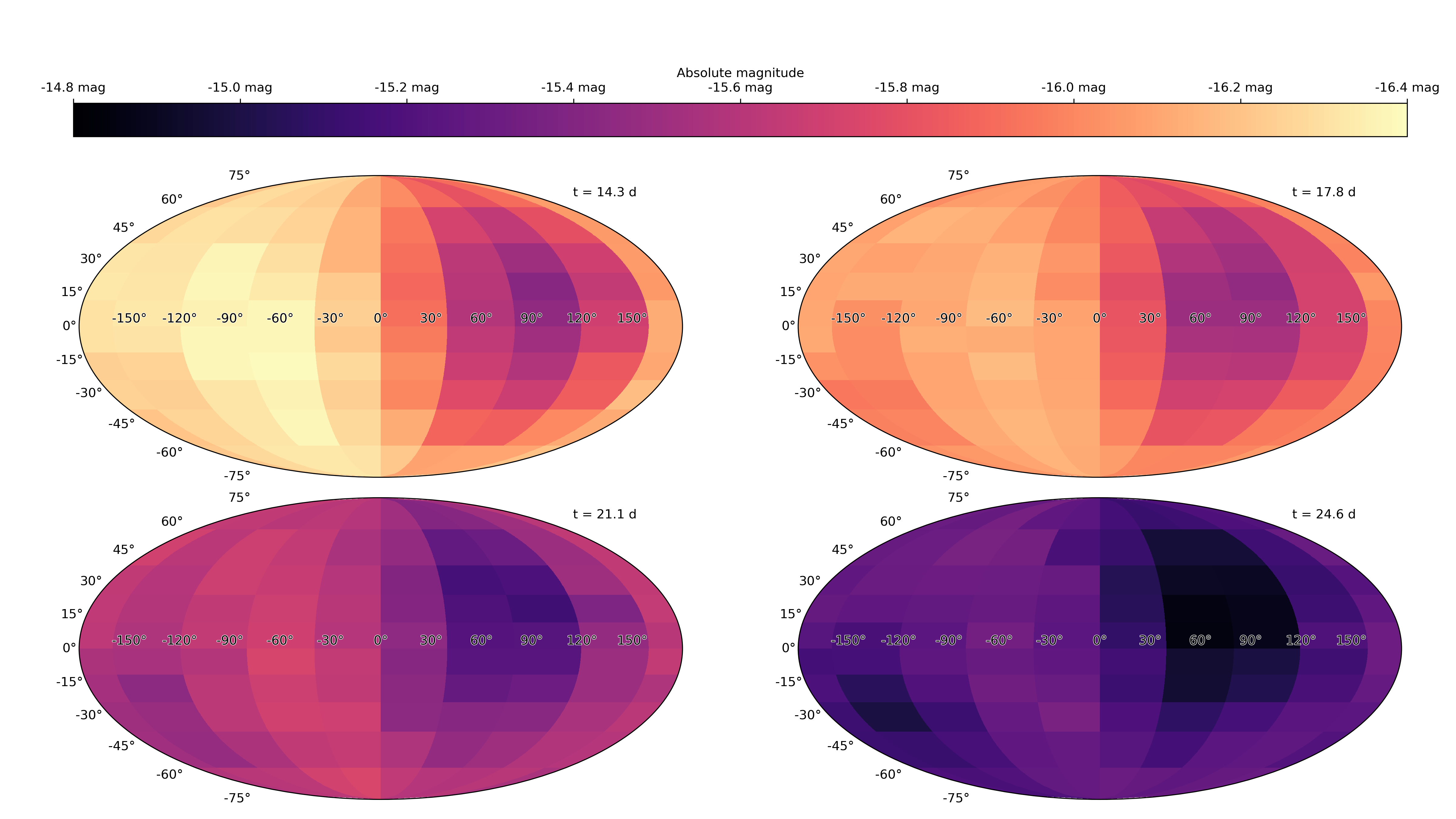}
    \caption{Viewing-angle dependence of the bolometric luminosity for various times from peak light (top left) to about 25 days after the onset of the explosion. Times (in days) are indicated in days next to the luminosity maps. }
    \label{fig:multi_time}
\end{figure*}

Dipolar variations in peak-magnitude with viewing angle
are present, but
not perfectly uniform across the entire spectrum.
Figure~\ref{fig:multiplot_plight} shows the direction-dependent magnitudes at peak light for the bolometric and U, B, V, R, and I bands. 
The variation in peak magnitude is most pronounced
in U band with a variation of $1.24\,\mathrm{mag}$ and less pronounced at longer wavelengths like the R-band with a variation of $0.79\,\mathrm{mag}$. It is strongest
up to and around peak light, and then diminishes
later on (Figure~\ref{fig:multi_time}).

\begin{figure}
    \centering
    \includegraphics[width=\linewidth]{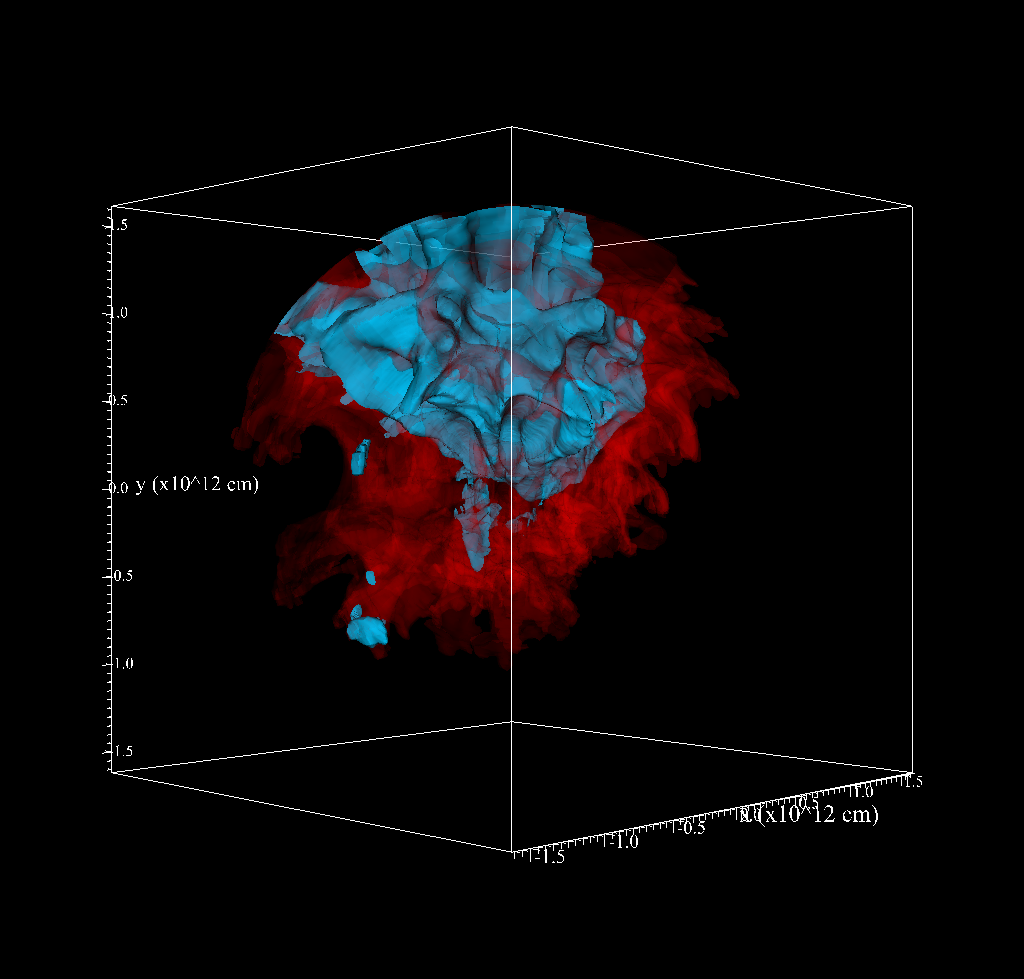}
    \caption{3D contour plot of the distribution of helium (red) and iron group elements (blue) at time of mapping to \artis. The contour levels for the mass fractions of }helium and the iron group elements are 0.25 and 0.15, respectively. While the overall ejecta are already somewhat asymmetric, the iron group elements are almost entirely contained within one hemisphere of the ejecta. Note that the structure shown in this plot is still significantly smaller than the entire star. At this time, the stellar surface has expanded to a radius of $9\times 10^{12}\,\mathrm{cm}$.
    \label{fig:3d}
\end{figure}

The viewing-angle dependence can be explained by the asymmetric distribution of $^{56}\mathrm{Ni}$ in the explosion model. This phenomenon has already been studied
in toy models of Type~Ia supernovae with parameterised global asymmetries in the $^{56}\mathrm{Ni}$ distribution
\citep{Sim2007}.
If the bulk of $^{56}\mathrm{Ni}$ is
displaced from the centre of the explosion, photons can more quickly diffuse and escape to the photosphere in the direction aligned with blob
because of the lower optical depth from the blob to the photosphere.
In modern 3D core-collapse supernova simulations, such large-scale asymmetries in the  $^{56}\mathrm{Ni}$ distribution emerge
naturally from the asymmetries in neutrino-driven convection during the early explosion phase, which
 usually become dominated by dipole, and sometimes by quadrupole modes around the time of shock revival \citep{mueller_19a}. 
 In stripped-envelope supernovae, these initial
 large-scale asymmetries are well preserved until
 and beyond shock breakout due
 to limited additional small- and medium-scale mixing
 by the Rayleigh-Taylor instability \citep{Chevalier1976}.
 Figure~\ref{fig:3d} shows the distribution of 
 iron-group elements (as a proxy for $^{56}\mathrm{Ni}$) around the time of mapping to \artis.

\begin{figure}
    \centering
    \includegraphics[width=\linewidth, trim={0.3cm 0.6cm 0.3cm 0cm}, clip]{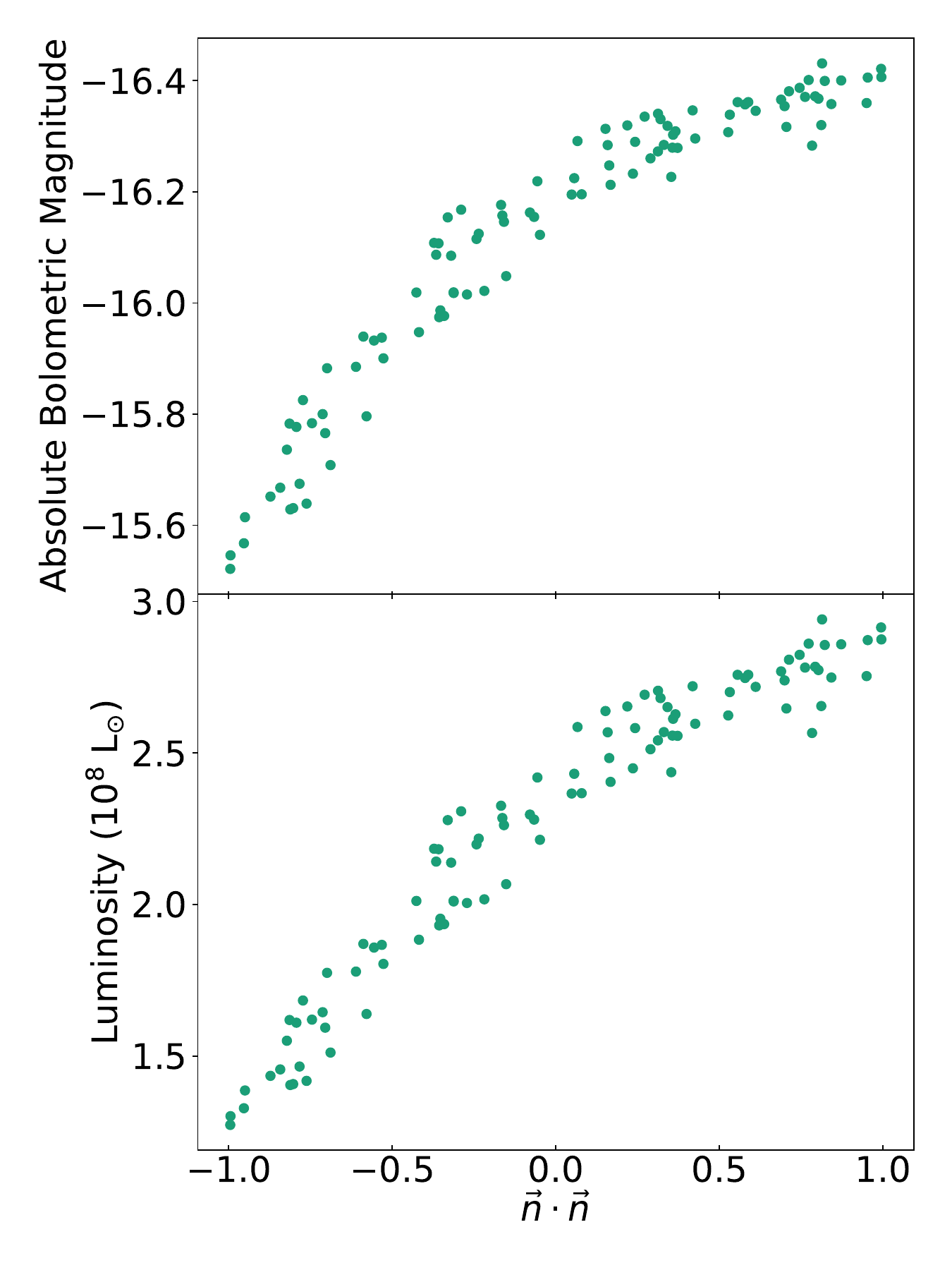}
    \caption{Peak bolometric luminosity (top panel: in magnitudes, bottom panel: on a linear scale)
    as a function of the angle
    cosine $\mathbf{n}\cdot \mathbf{n}'$ between
    the observer direction $\mathbf{n}'$
    and the direction $\mathbf{n}$ of the centre of mass  of the iron-group ejecta.
    The luminosity is highest for observers in the same direction as the iron-group blob.
    Note the near linearity in luminosity with respect to $\vec{n}\cdot\vec{n}$. }
    \label{fig:distnorm}
\end{figure}

To illustrate that the viewing-angle dependence is aligned 
with the asymmetries in the iron-group ejecta,
Figure~\ref{fig:distnorm} shows the absolute bolometric
magnitude and luminosity in solar units for all observer directions $\mathbf{n}'$
as a function of the squared distance $|\mathbf{n}-\mathbf{n}'|^2
=1-2 \mathbf{n}\cdot\mathbf{n}'$
to the  centre-of-mass of the iron group elements
$\mathbf{n}$. There is a clear correlation between the brightest areas of the photosphere and the (non-central) location of the bulk of the iron-group ejecta.  The  dependence of the bolometric magnitude on  $\cos \theta$ is reasonably tight.
The dependence of the bolometric luminosity
on $\mathbf{n}\cdot\mathbf{n}'$ is fairly linear, so the \emph{average} bolometric luminosity
for a random observer will not be strongly
biased away from the mean value.
The distribution of observed magnitudes will be skewed, however, simply because magnitudes constitute a logarithmic scale.
Over about $75\%$ of the sphere, the peak absolute
magnitude varies only by about $0.5\,\mathrm{mag}$. The probability
to observe the supernova as significantly dimmer than in the
brightest direction (e.g., by more than $0.5\,\mathrm{mag}$)
is relatively small for a randomly located observer.

\begin{figure}
    \centering
    \includegraphics[width=\linewidth]{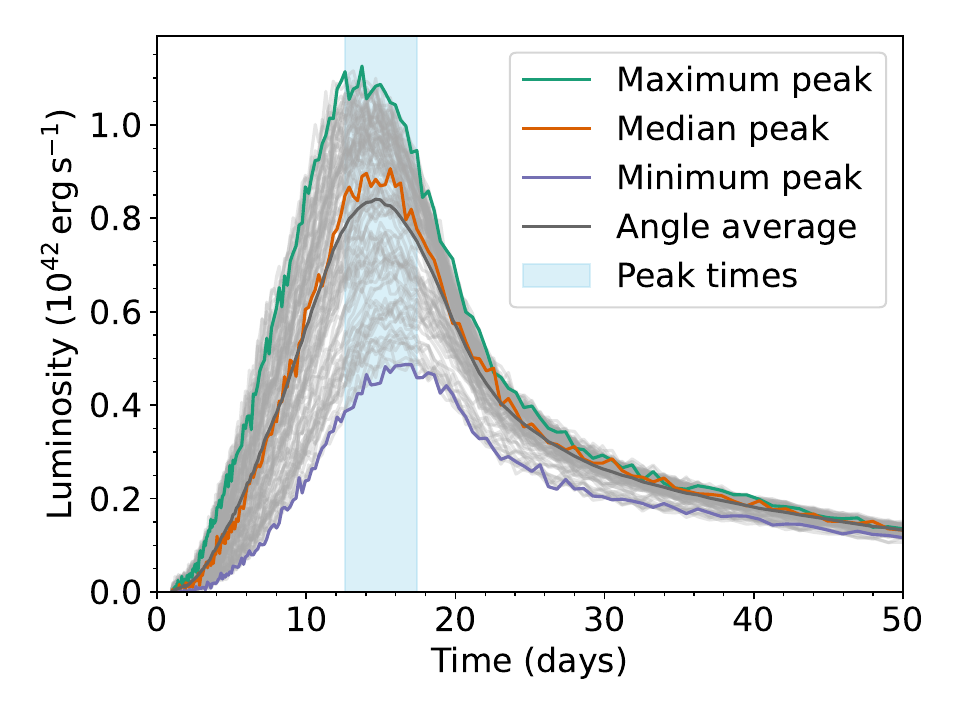}
    \caption{Viewing-angle dependent bolometric light curves for our baseline model. The light curves with the brightest and dimmest peak, and the
    light curve corresponding to the median of the distribution
    are shown in green, dark purple, and orange. The angle-averaged
    light curve is also shown.}   
    The light blue region denotes the range of peak times for all different viewing angles.
    
    \label{fig:bounds}
\end{figure}

To illustrate the variations in the light curves for different observer directions,
light curves for all 100 viewing angles are shown in grey in Figure~\ref{fig:bounds}, together with the light curves corresponding to the minimum, median, and maximum peak brightness.
Figure~\ref{fig:bounds} shows that
the viewing-angle dependence does not only
affect peak luminosity, but also concerns the
shape of the light curve. In those directions where the maximum luminosity is higher, the light curve is also more strongly peaked as shown in Figure~\ref{fig:bounds}, with a \emph{faster}
evolution of the light curve around maximum brightness.  The viewing-angle dependence is most
pronounced before and around peak and diminishes during the later phases. In addition, the time of peak also varies slightly; the range of peak epochs is indicated as the light blue region in Figure~\ref{fig:bounds}.
Such a strong viewing-angle dependence in the light curve has important implications for comparisons to observed stripped-envelope supernovae. This will be discussed in
Section~\ref{sec:pop_comp}.

\section{Spectroscopy}
\begin{figure}
    \centering
    \includegraphics[width=\linewidth]{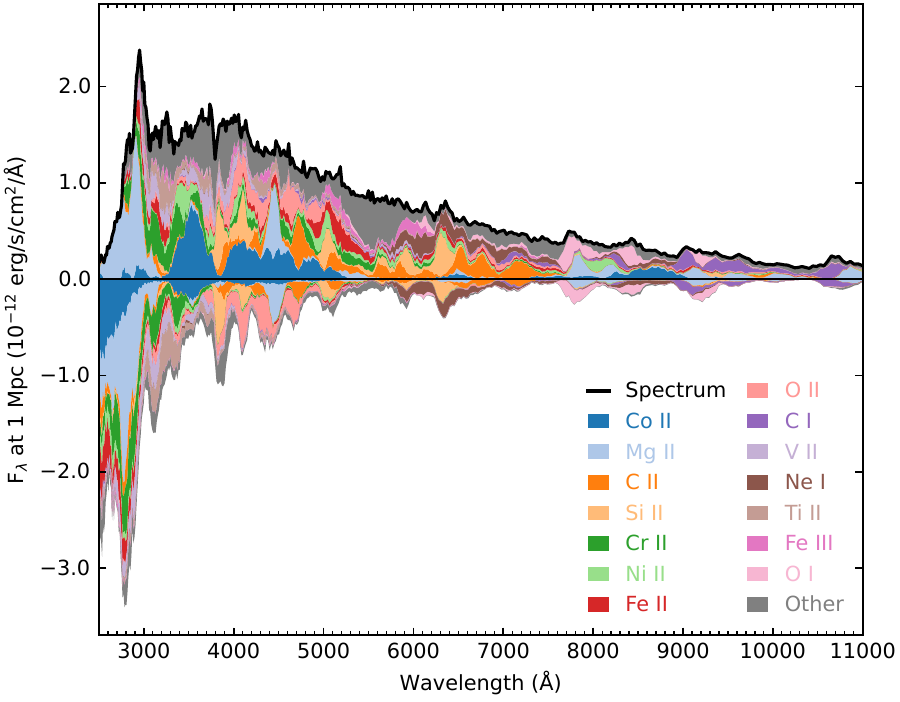}
    \caption{
    Synthetic spectrum at peak light for our baseline model. 
    Contributions of the 14 most abundant sources of emission (top halves of panels) and absorption (bottom halves) are shown as stacked plot; colours indicate the ion responsible for the last interaction. There is a prominent Mg~II feature at 2,800 \AA. Overall, this early spectrum shows few identifiable features, similar to the more extreme ultra-stripped supernovae model of \citet{Maunder2024}. The spectrum is largely dominated by iron-group elements, Mg~II, and some emission from silicon, carbon, oxygen and calcium that only contribute minor features.
    }
    \label{fig:specp}
\end{figure}

\begin{figure}
    \centering
    \includegraphics[width=\linewidth]{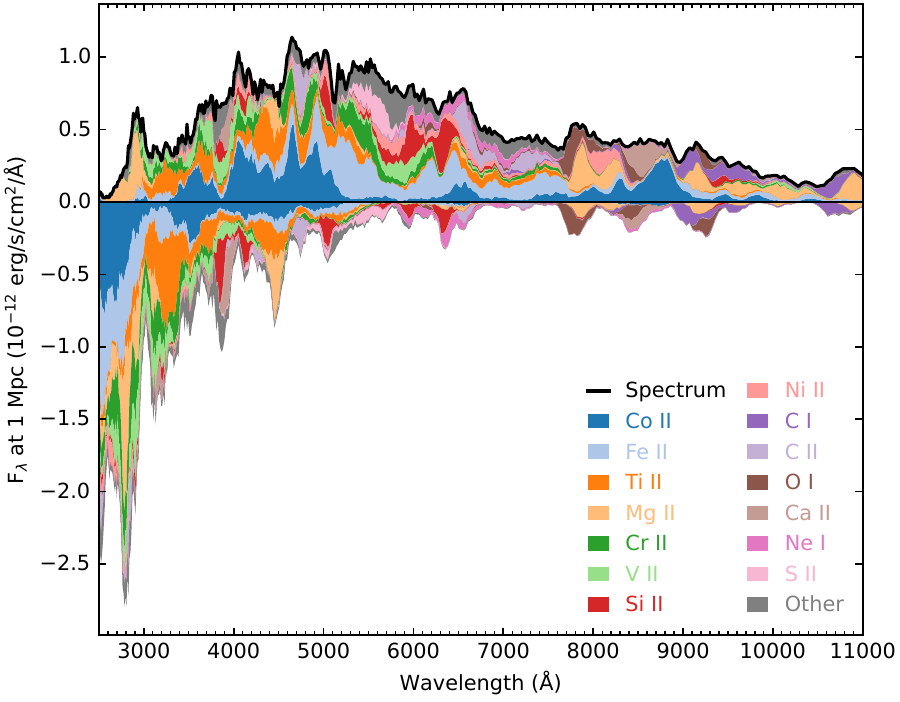}
    \caption{Synthetic spectrum around $6\,\mathrm{d}$ after peak for the baseline model.
    The prominent Mg~II feature at $2\mathord,800\,\text{\AA}$ is still present. C~II and and Si~II features are becoming more clearly visible, and the 
    \wl{7}{774} O~I line appears, blended
    with a  Mg~II feature.
    }
    \label{fig:spec+6}
\end{figure}

\begin{figure}
    \centering
    \includegraphics[width=\linewidth]{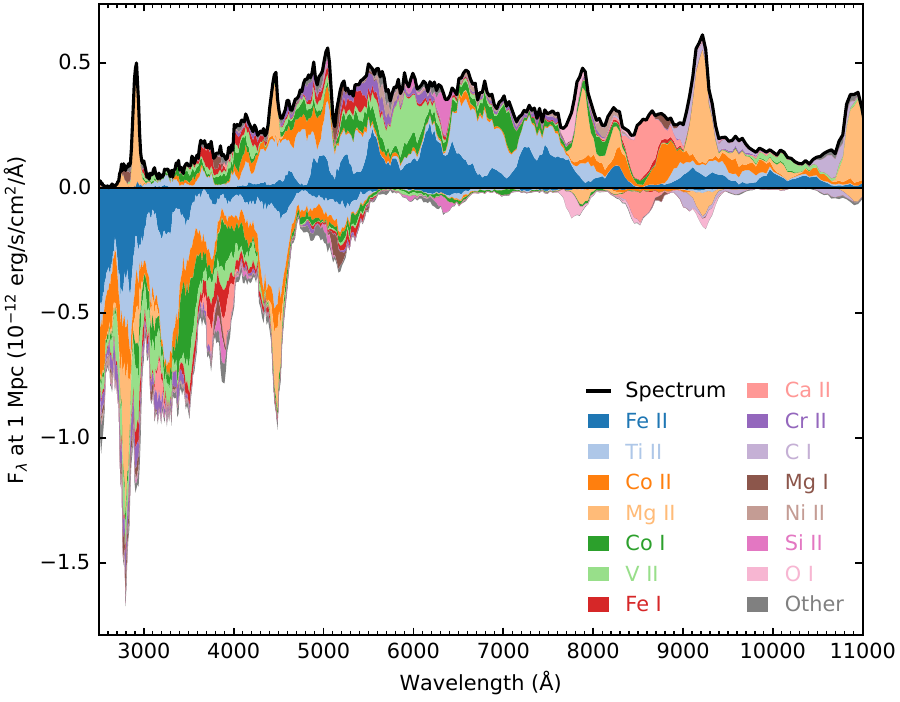}
    \caption{
    Synthetic spectrum around $12\,\mathrm{d}$ after peak for the baseline model.
    The Mg~II feature at 2800 \AA\ is becoming more prominent again, and more Mg~II
    lines appear in the infrared, as well as the Ca NIR triplet. Identifiable C~II and Si~II
    features are no longer present.
    }
    \label{fig:spec+14}
\end{figure}

Along with photometric information, we also present  synthetic spectra for our model. Figures~\ref{fig:specp}, \ref{fig:spec+6}, and \ref{fig:spec+14} 
show the emerging angle-averaged spectrum 
(upper panels) at peak light, $6\,\mathrm{d}$ post-peak and $12\,\mathrm{d}$
post-peak, along with a breakdown into the photon interactions that shape the spectrum.
The emerging spectrum in the top halves of panels is broken down into the contributions of the 14 most abundant sources of emission that created the photon packets, with colours indicating the ion or process involved in the last interaction.
To obtain an angle-averaged spectrum, we construct the spectrum from emerging Monte Carlo packets in any direction, as with the angle-averaged
light curves.
The bottom halves of the panels show the flux distribution before the last interaction event, which provides an indication for strong scattering or absorption from these wavelength regions.

The most notable feature in the spectra is the prominent Mg~II peak at \wl{2}{800}, the same strong line found in the more extreme ultra-stripped supernova model used in our previous work in 2D \citep{Maunder2024}. Aside from that, the spectra do not yet show other clearly identifiable P Cygni lines as one would normally expect from Type Ib/c's, but there are some features that can be distinctly associated with specific ions.
Compared to the 2D model of \citet{Maunder2024}, some
familiar lines from Type Ib/c spectra \emph{do} appear in this
new model with higher ejecta mass, even though they still
differ in detail from observed events.
There is another weak Mg~II feature in the optical at \wl{4}{600} which becomes prominent at later times after peak, and some other Mg features in the infrared also become more prominent with time. As found in our previous work, much of the emission comes from iron group elements, namely vanadium, titanium, chromium, iron, and cobalt. More interesting is the presence of C~I, C~II, Si~II and O~I lines, which had not appeared as strongly or as early in \citet{Maunder2024}, or had not appeared at all. 

There is a noticeable
\wl{3}{858}  Si~II line already at the time of peak.
Si~II also forms a feature at around \wl{6}{300} which then disappears by 14 days post-peak. This line is a commonly observed feature in Ic supernovae \citep{Holmbo2023}, although the feature appears only transiently in our model. In the spectrum
at 6 days, Si~II contributes to several features in the optical
and near UV.

The \wl{7}{774} O~I line, which is commonly
observed in stripped-envelope supernovae
\citep[e.g.,][]{fremling_18} is already present at peak
and remains quite prominent at 6 days after peak, albeit
in a blend with a Mg~II feature. At late times, the contribution
of O~I to the blended features is still present, but becomes less prominent.
One of the most commonly observed emission features in type Ib/c supernovae, the [O~I] \wl{6}{300} and \wl{6}{363} lines, are still absent in our models. This is simply due to the fact that these forbidden transition
of oxygen are not included in the atomic dataset used in this study
(cp.\ also the discussion of atomic processes relevant
to oxygen line formation in \citealt{van_baal_23}).
More generally, the approximate treatment of NLTE in \textsc{Artis} as adopted in this study, will affect the formation of some lines, and is a particularly important limitation for late-time, nebular-phase spectra.

After peak, C~II contributes to a narrow feature at around \wl{4}{800} and a broad feature around \wl{6}{600}-\wl{6}{800}, which both disappear by 12 days post-peak. 
At late times the Ca~II triplet around \wl{8}{500} was a prominent broad feature in our previous work, which is now obscured somewhat by the large Mg~II feature around \wl{9}{400}. There is another Mg~II peak around \wl{10}{900}. These were not seen to such an extent in our previous work, 
but the
Mg~II 
P Cygni
feature at \wl{2}{800} is less prominent than in the more
extremely stripped-model from \citet{Maunder2024}. The
presence of additional Mg~II features is understandable
in the light of the higher \emph{total} mass of
Mg in the current model.
Compared to more typical stripped-envelope supernovae the absolute mass of ejected Mg of $0.021\,\mathrm{M}_\odot$ of magnesium and the overall Mg mass fraction
are not unusual for progenitors with a low C/O core mass, though smaller than expected from supernova
explosions of stars with more massive cores. Extrapolating the predicted shape of the spectrum 
and the presence and strength of Mg lines
to progenitors with more massive C/O cores would be premature. In the explosions of more typical stripped-envelope supernovae, the Mg will be surrounded by more massive He envelopes and more
massive shells of C and O, and the mixing will also proceed differently.

\begin{figure}
    \centering
    \includegraphics[width=\linewidth]{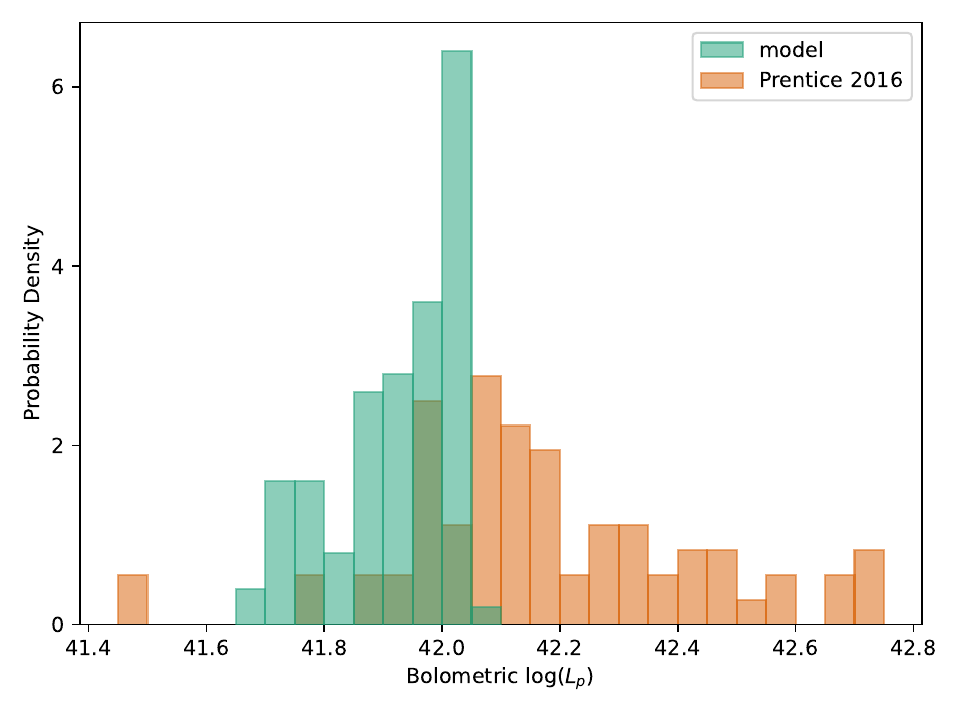}
    \caption{Histogram (sample estimate of the probability density)} of peak bolometric magnitude across viewing angles in
    our model and for the sample of observed Type~Ib/c supernovae of \citet{Prentice2016}. 
    \label{fig:mag_hist_model}
\end{figure}

\begin{figure}
    \centering
    \includegraphics[width=\linewidth, 
    clip]{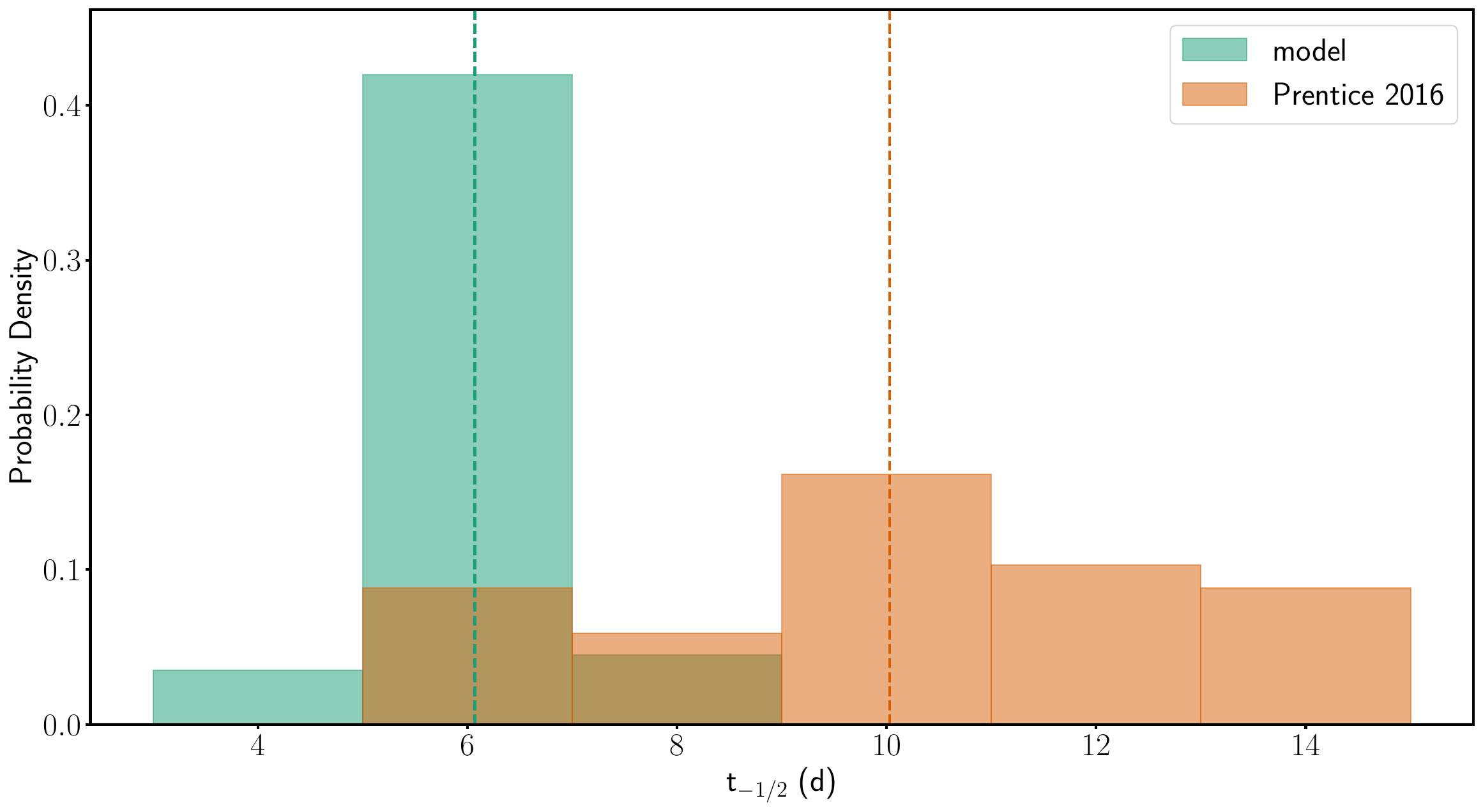}
    \centering
    \includegraphics[width=\linewidth, 
    clip]{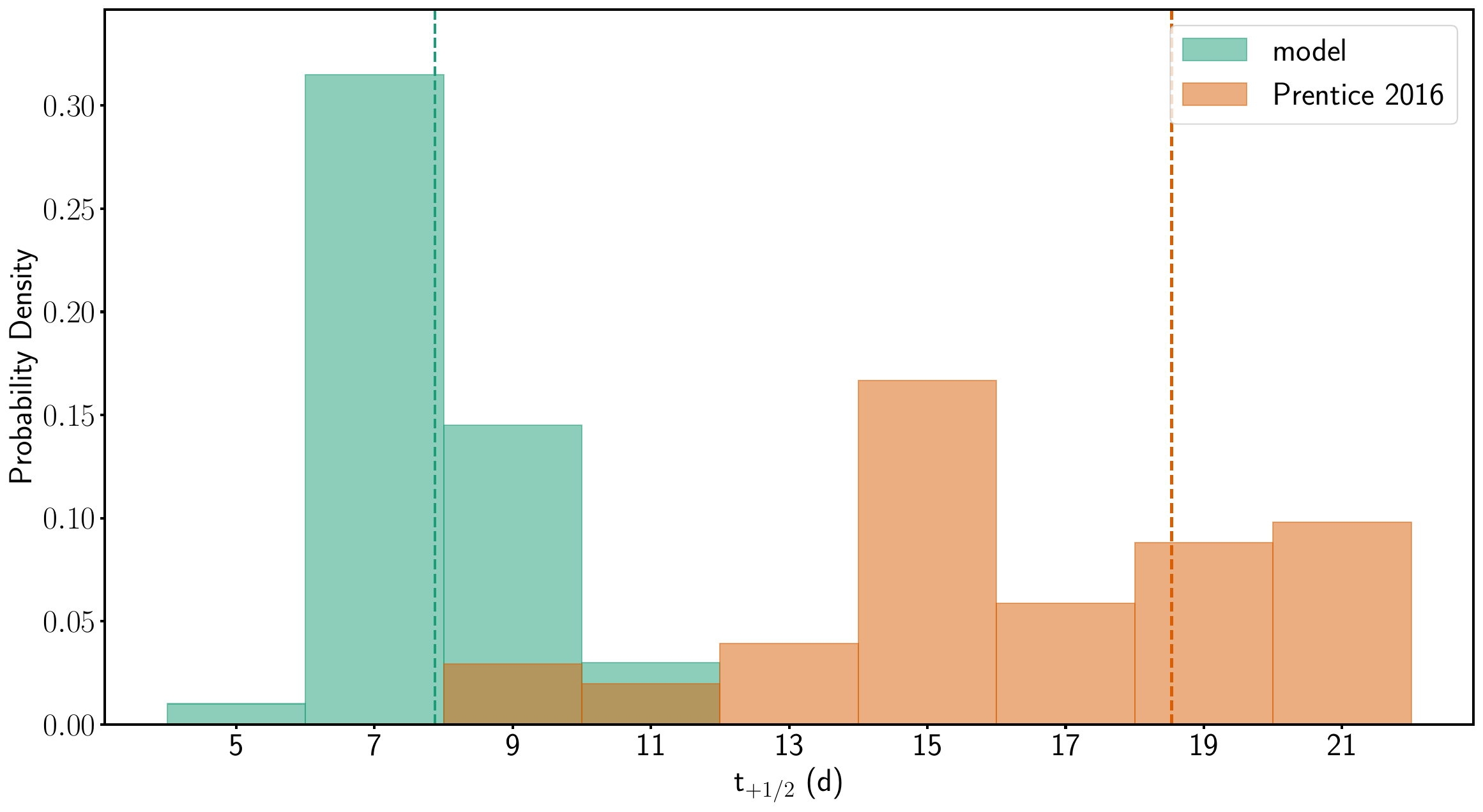}
    \caption{Comparison of the distribution of
    the half-rise time $t_{-1/2}$ and
    the half-decay time $t_{+1/2}$ across
    viewing angles with the observed distribution of  $t_{+1/2}$ in the observed Ib/c supernova sample of \citet{Prentice2016}.
    Green and red vertical lines show the mean values for the model across viewing angles
    and for the sample from \citet{Prentice2016}, respectively.}
    \label{fig:obs_comp_decaytime}
\end{figure}

\begin{figure}
    \centering
    \includegraphics[width=\linewidth]{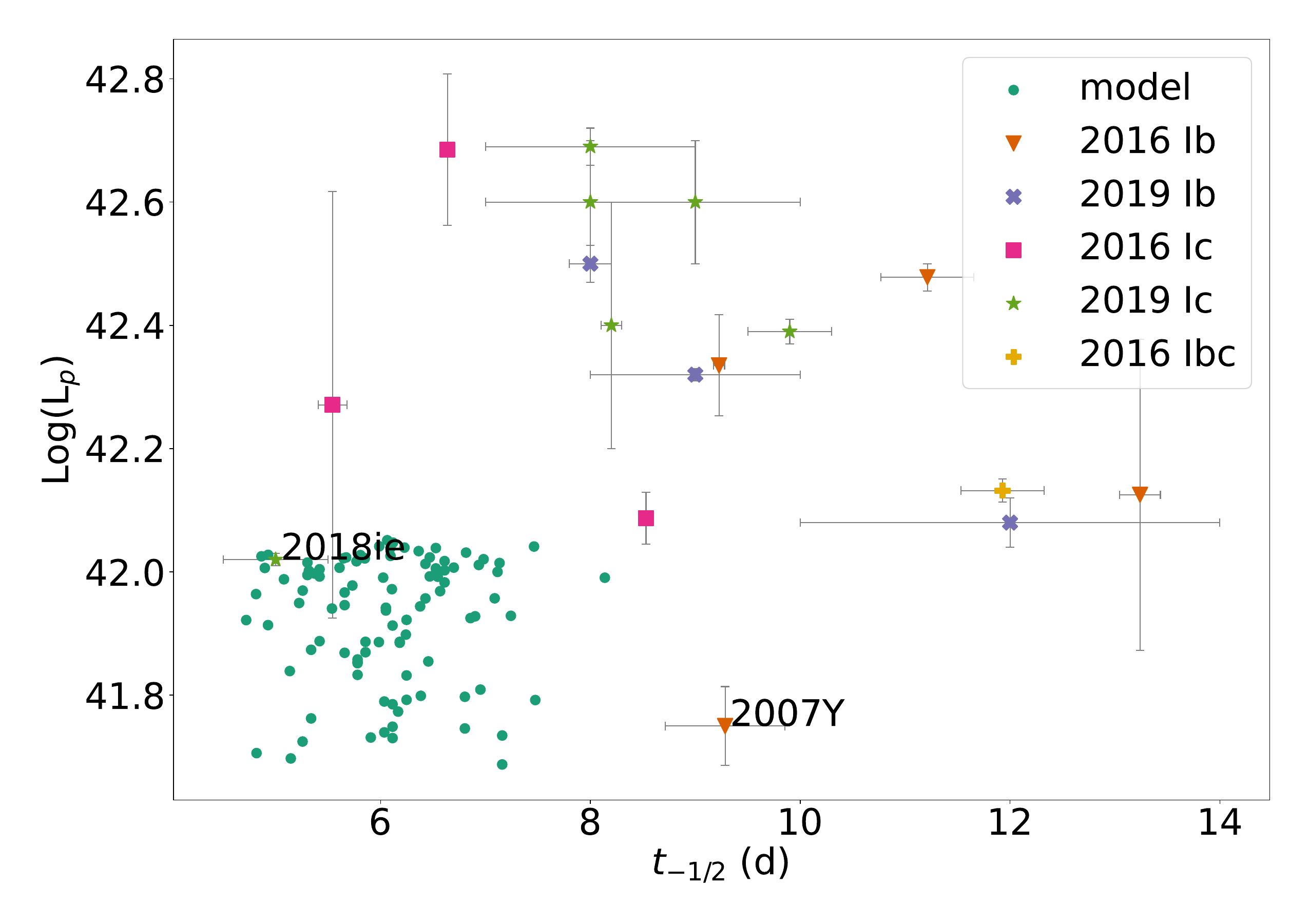}\\
    \includegraphics[width=\linewidth]{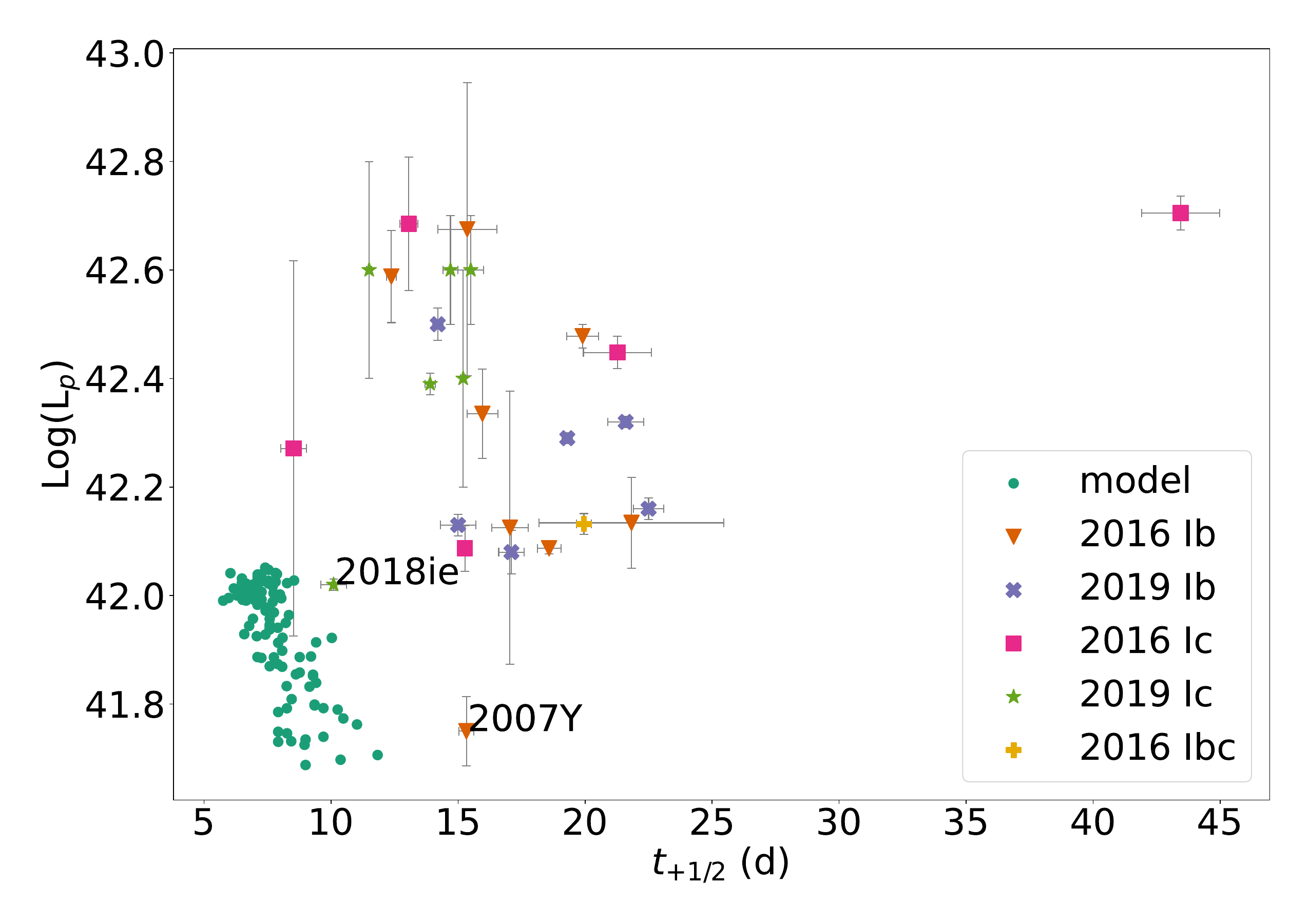}
    \caption{Distribution of peak bolometric luminosity $L_\mathrm{p}$
    versus  half-rise time $t_{-1/2}$ (top panel)
    and
    half-decay time $t_{+1/2}$ (bottom panel) across viewing angles 
    for our model, and for the stripped-envelope supernova samples from \citet[][labelled as ``2016'']{Prentice2016} and \citet[][labelled as ``2019'']{Prentice2019}.
     Again the observational samples are broken down into
    Ib and Ic supernovae , with the addition of some unclear cases (Ibc). Only transients with
    observed data for both $t_{-1/2}$ \emph{and} $t_{+1/2}$ are shown for consistency between the two plots. Broad-lined Ic (Ic-BL) supernovae and IIb supernovae from \citet{Prentice2016} are not shown.}
    \label{fig:prentice_rise_decay}
\end{figure}

\begin{figure}
    \centering
    \includegraphics[width=\linewidth]{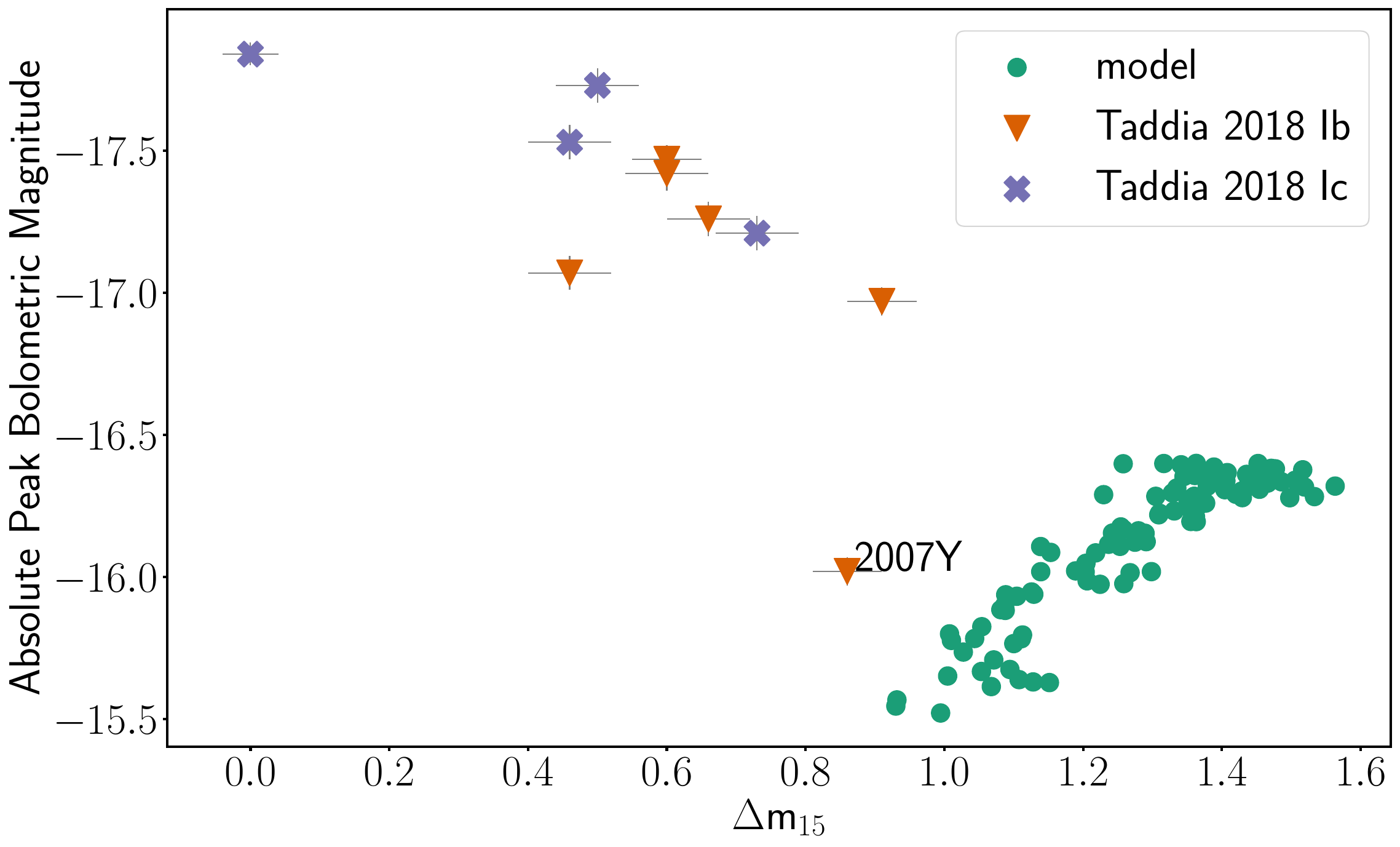}
    \caption{
    Distribution of peak bolometric magnitude versus
        $\Delta$m$_{15,\mathrm{V}}$ across viewing angles for our model (circles),
        and for the stripped-envelope supernova sample of  \citet[][symbols with error bars]{Taddia2018}.
The observed sample is  broken down into
    Ib and Ic supernovae using different
    colours and symbols. 
    Again, Ic-BL and IIb supernovae are not shown. Note that peak bolometric luminosity \emph{and} $\Delta m_{15}$ are only available for a subset of the Taddia sample.}
    
    \label{fig:dm15bolo}
\end{figure}

\begin{figure}
    \centering
    \includegraphics[width=\linewidth]{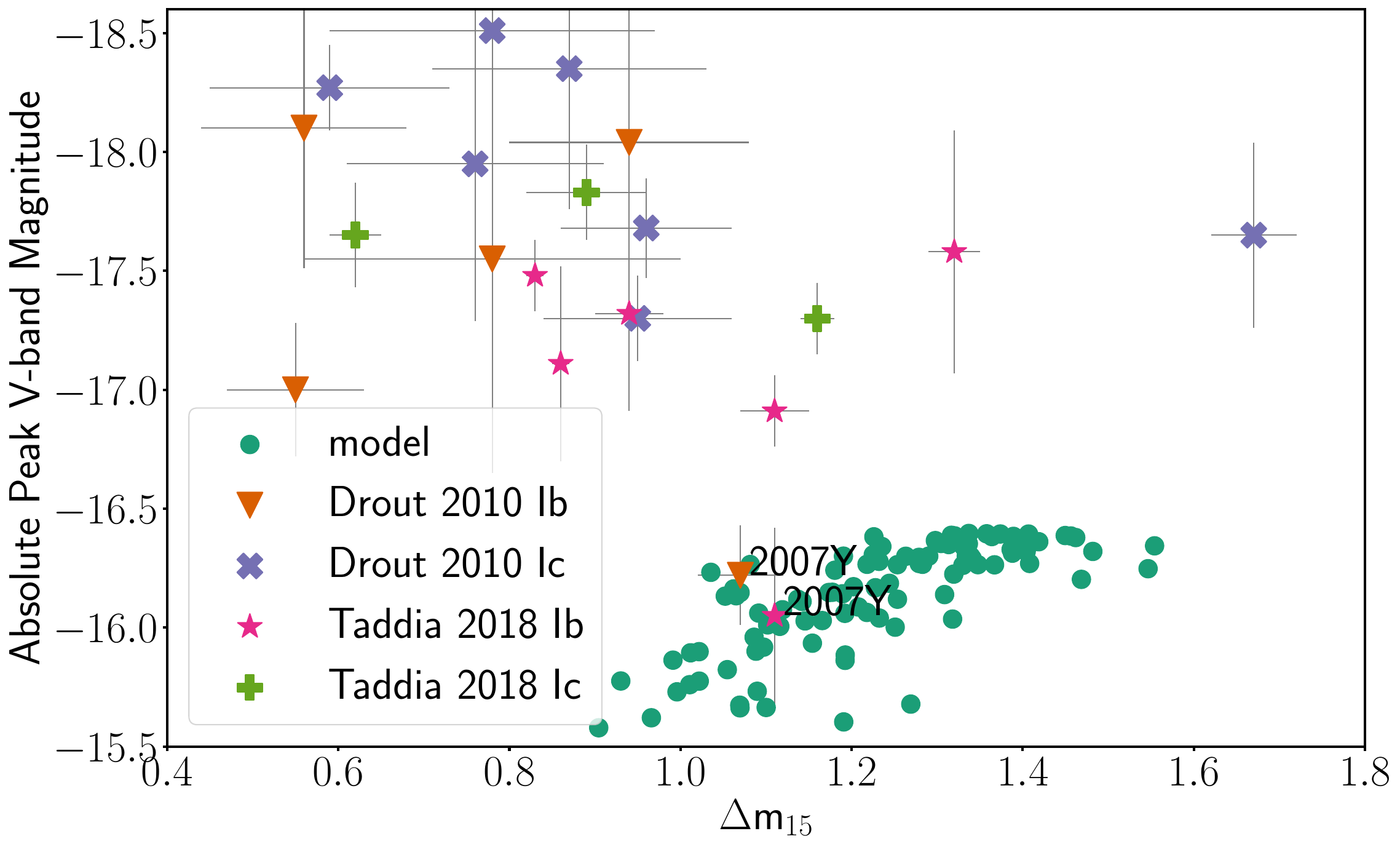}
    \caption{Distribution of peak absolute V-band magnitude versus
        $\Delta$m$_{15,\mathrm{V}}$ across viewing angles for our model (circles),
    and for the stripped-envelope supernova samples (symbols with error bars) of \citet{Taddia2018} and \citet{Drout2011}. The observational samples are further broken down into
    Ib and Ic supernovae using different
    colours and symbols.
    Again, Ic-BL and IIb supernovae are not shown. Note that peak V-band luminosity \emph{and} $\Delta m_{15}$ are only available for a subset of the samples.}
    
    \label{fig:dm15V}
\end{figure}

\section{Comparison with the Observed Type~Ib/c Supernova Population}
\label{sec:pop_comp}
In comparing to observed stripped-envelope supernovae, we limit
ourselves to key features of the photometry since the synthetic spectra still exhibit characteristics very distinct from observed events. Because of the strong viewing-angle dependence of the predicted photometry, it is most appropriate to compare the distribution of photometric properties for different observer directions with the observed population of stripped-envelope supernovae, rather than to focus unduly on matching individual events. 

We compare our model to various observational studies of the supernova population, focusing on the Type Ib/c events. The surveys cover a variety of parameters to measure the evolution timescales of the explosions across different optical bands. The surveys from \citet{Prentice2016, Prentice2019} measure the evolution timescales using the time taken to rise from half of the peak luminosity to peak, $t_{-1/2}$, and the time taken to decay to half of the peak luminosity from time of peak, $t_{+1/2}$. Here we use their bolometric luminosities. The survey in \citet{Taddia2018} provides $\Delta\mathrm{m}_{15}$ values, which measures the change in magnitude over the first 15 days post-peak light. Here we use the V-band and bolometric values. Finally, we use the survey of \citet{Drout2011} who also provide $\Delta\mathrm{m}_{15}$ values and here we use their V-band results. Note that these are not controlled, volume-limited surveys and may hence suffer from a bias towards brighter events.

The comparison to observed transients still places our model at the
fast and faint end of the observed stripped-envelope supernova population.  Different from the more extreme ultra-stripped supernova
model in \citet{Maunder2024}, the distribution of luminosity across
observer directions overlaps with the observed luminosity distribution, however. Comparing to the sample of \citet{Prentice2016}, the distribution of peak bolometric luminosity
overlaps well with the lower half of their observed distribution and
reaches almost up to the mean value of their sample thanks to the enhanced brightness in the direction of the iron-group blobs
(Figure~\ref{fig:mag_hist_model}).
However, our model evolves significantly faster than most observed transients, as can be seen from
histograms of the half-rise $t_{-1/2}$ and half-decay time $t_{+1/2}$
(Figure~\ref{fig:obs_comp_decaytime})
for the distribution across observer directions in comparison to
the samples of \citet{Prentice2016,Prentice2019}.

We highlight two events that sit within our model's regime in at least some bands and evolution metrics. Firstly, SN~2007Y is a Type Ib supernova of similar brightness to our event, however, as can be seen in Figures~\ref{fig:prentice_rise_decay}-\ref{fig:dm15V}, the time evolution does not match our model.
SN~2007Y is seen to evolve slower both in the rise time and decay time, supported further by the lower $\Delta m_{15}$ in both the V- and R-band. For a direct comparison, we also show the observed B- and V-band light curve for SN~2007Y in Figure~\ref{fig:cband_comp}. The figure illustrates the somewhat broader peak of SN~2007Y compared to the angle-averaged model light curve. The decline of SN~2007Y does not exhibit the pronounced bump or step-like structure around 20 days that is present in the model. There is a hint of a non-steady decline in the B-band light curve of SN~2007Y around 30 days, but this is much less conspicuous than in the model.

SN~2018ie sits within our distribution of models in the bolometric band half-rise time from \citet{Prentice2016, Prentice2019} and is missing from the sample presented by \citet{Drout2011} and \citet{Taddia2018}. This event is
also within the range of half-decay times for our models,
but these slowly decaying models are dimmer than  SN~2018ie.
Our model is fainter than the observed sample except for SN~2007Y
Our model decays more rapidly than the observed sample aside from SN~2007Y which has a comparable decay time from peak to half-peak.
SN~2021gno \citep{ertini_23} is another stripped-envelope supernovae with comparably low luminosity. However, SN~2021gno can be ruled out as candidate for a closer comparison because of the slower decay, and also has a double-peaked
light curve that would require interaction with
circumstellar material, which is not present in our model, but may occur
for more extreme ultra-stripped supernova progenitors in which degenerate burning can
result in pre-supernova mass ejection
and produce an interaction-powered transient
\citep{moriya2025}.

This mismatch with the observed distribution may simply indicate that the simulated explosion of an ultra-stripped supernova is still too rare an event to be covered by the observed samples due to selection biases. \citet{Tauris2013} estimate the fraction of ultra-stripped Type Ic supernovae to the total supernova rate to be in the range of $10^{-2}-10^{-3}$. Sampling biases in the observed distribution clearly play some role; e.g., there is much less overlap of our predicted  viewing-angle dependent brightness distribution with the  peak V-band magnitude
of \citet{Drout2011,Taddia2018} and the peak bolometric luminosities of \citet{Prentice2016,Prentice2019}. When considering 
the rise or decay time together with peak luminosity, it is even clearer
that our simulation occupies an hitherto unobserved niche in 
parameter space (Figure~\ref{fig:prentice_rise_decay}).
Considering the low expected events rates, the observed samples (and the number off well-observed faint events) simply remain too small to rule out models for ultra-stripped supernovae
or to formulate criteria for classifying explosions from ultra-stripped progenitors and place rate limits on them.

However, the predicted viewing-angle
dependence and observed or non-observed correlations of light curve properties may provide a path towards constraining supernova explosion dynamics. Our models show a strong correlation between 
peak bolometric brightness and  $\Delta\mathrm{m}_{15}$, i.e.,
the decay after peak is faster for higher peak luminosity (Figure~\ref{fig:dm15bolo}). The spread in $\Delta\mathrm{m}_{15}$ associated with this correlation amounts to almost $1\,\mathrm{mag}$.
To some extent this correlation holds for the peak V-band magnitude as well
(figure~\ref{fig:dm15V}).

The observed population does not show this
correlation between brightness and $\Delta\mathrm{m}_{15}$; 
there rather is an anti-correlation with brighter events evolving more slowly. This is especially
clear when considering bolometric luminosity
(Figure~\ref{fig:dm15bolo}), where the observed
width of the distribution of peak luminosity
versus $\Delta\mathrm{m}_{15}$ is quite
narrow in the set of \citet{Taddia2018}.

This has important implications for the degree of
asymmetries in nickel distribution in stripped-envelope supernovae. The overall distribution of and correlations between peak brightness and $\Delta\mathrm{m}_{15}$ will result from the combination of the viewing-angle variations
and progenitor-dependent variations resulting from different
ejecta masses, nickel production, and explosion energies.
Thus, if the asymmetries in the nickel distribution and the peak luminosity were of similar magnitude as in our model
across \emph{all} stripped-envelope supernovae,
the observed distribution of $\Delta\mathrm{m}_{15}$ versus peak luminosity would have to reflect this even after the viewing-angle distribution is convolved with progenitor-dependent variations. Progenitor variations could still give rise to the observed anticorrelation of $\Delta\mathrm{m}_{15}$
versus peak luminosity, but it would be impossible
to obtain the narrow distribution of $\Delta\mathrm{m}_{15}$ if there was the same amount of intrinsic scatter of $\Delta\mathrm{m}_{15}$ for individual explosions as in our model. Even if we disregard that
the scatter in our model goes in the wrong
direction (bigger $\Delta\mathrm{m}_{15}$ 
for brighter events), the spread in $\Delta\mathrm{m}_{15}$ of $\mathord{\sim}0.7\,\mathrm{mag}$
across viewing angles for a \emph{single} progenitor is already comparable to that of the observed sample, leaving little room for extra variance.
In the direction of the predicted anticorrelation, the scatter of the models is arguably larger than
that in the sample  of \citet{Taddia2018} already.

Thus,  the asymmetries in our model may be too large to be representative of \emph{generic} stripped-envelope supernovae. This does not yet conclusively prove that current 3D supernova
explosion models are too asymmetric. 
The transient set from  \citet{Taddia2018} used in 
Figure~\ref{fig:dm15bolo} is small, so a quantitative comparison of the scatter would not be enough to permit
firm conclusions. 
There seems to be less of a tension when comparing to he 
bigger sample of Ib/c supernovae with measured peak V-band magnitude  in Figure~\ref{fig:dm15V}.
Furthermore, in
Ib/c supernovae with more massive envelopes of helium and heavier elements, the final distribution of nickel may be less asymmetric, and its impact on the viewing-angle dependence of the light curve may be mitigated by the large optical depth of these outer layers. However, our results demonstrate that the predicted viewing-angle dependence provides a means to constrain explosion asymmetries based on the observed distribution of light curve properties, if either the observed samples can be extended to lower magnitude, or simulations for more massive progenitors become available.
Interestingly, the amplitude of the viewing-angle dependence
of about $1\,\mathrm{mag}$ at peak is even larger than in the
artificial jet-driven explosion models of \citet[][Figure~2]{rapoport_12}, except perhaps for the U-band, where
\citet{rapoport_12} find a strong direction-dependence during
the rise phase. The less pronounced viewing-angle dependence in
the jet-driven model could be related to its considerably
higher ejecta mass of $10\,\mathrm{M}_\odot$. In future, it
will be important to investigate whether higher optical depth
from the nickel blobs and slower diffusion from the
nickel blobs to the photosphere reduce the viewing-angle dependence
in less extremely stripped neutrino-driven supernovae.

\section{Conclusions}
We performed 3D Monte Carlo radiative transfer calculations for a self-consistent explosion model of an ultra-stripped supernovae with a pre-collapse mass
of $2.4\,\mathrm{M}_\odot$.

The radiative transfer calculations show a relatively faint and rapidly evolving transient with
$\Delta m_{15} \sim \texttt{1-2}\,\mathrm{mag}$. Interestingly, we find a considerable viewing-angle dependence in the predicted photometry. The peak magnitude differs by about
$1\,\mathrm{mag}$ depending on the observer direction, ranging from $-16.4\,\mathrm{mag}$ down to $-15.3\,\mathrm{mag}$ in bolometric magnitude. The strong viewing-angle dependence is most pronounced up to peak and subsequently decreases. It is caused by the asymmetric distribution of $^{56}\mathrm{Ni}$ in the explosion. In directions aligned with the dominant iron-group blob, faster diffusion towards the photosphere results in a significantly enhanced luminosity. The asymmetric distribution of iron-group ejecta also affects the shape of the light curve. Observers in the direction of the iron-group blob would also see more rapidly evolving light curves with a faster rise and decline. This is similar to a phenomenon predicted for
toy models of Type~Ia supernovae with  asymmetric iron distribution \citep{Sim2007}.

Similar to the case of a more extreme ultra-stripped supernova of \citet{Maunder2024}, the predicted spectra are unusual for Ib/c supernovae and do not match any observed events. Like in 
 \citet{Maunder2024}, the most prominent feature is the Mg~II line at
  \wl{2}{800}.
  Mg~II lines also appear at \wl{4}{600} in the optical and at
 \wl{8}{000}, \wl{9}{400}, and \wl{10}{900} in the infrared during some phases.
Different from \citet{Maunder2024}, however, lines from other elements emerge as well.
The \wl{7}{774} O~I  line that is commonly found in Ib/c supernovae appears quite prominently in our model, although it
is blended with a Mg~II emission feature.
The forbidden O~I lines at \wl{6}{300} and \wl{6}{363}
are still absent, as these forbidden
transitions are not included in the atomic
data set used for this study.
We find C emission features around \wl{4}{800} and \wl{6}{600}-\wl{6}{800} as
well as Si~II feature around \wl{3}{858} and \wl{6}{300} during some epochs.
At late times the Ca~II NIR (near-infrared) triplet around \wl{8}{500} appears.
The spectral features in our model still differ in detail
from observed Ic supernovae -- too much to permit a meaningful
comparison with individual events -- but with the bigger ejecta
mass compared to the more extreme ultra-stripped model
of \citet{Maunder2024}, similarities with observed events
are emerging. 

Because of the strong viewing-angle dependence of the predicted photometry, it is appropriate to compare the range of predicted photometric properties to the distribution of observed Type~Ib/c supernovae rather than to focus unduly on a comparison to single events. Among better observed individual Ib/c supernovae, SN~2007Y is the only one to which our viewing-angle dependent light curves come close
to in terms of peak magnitude and rapid evolution.  The low luminosity relative to observed
Type~Ib/c events may still be exaggerated by selection biases. When compared to the sample
of \citet{Prentice2016}, our predicted viewing-angle dependent bolometric luminosities cover the
entire lower half of the observed sample. The comparison to the observed population of stripped-envelope
supernovae is more interesting and potentially constraining for supernova explosion models when considering correlation between peak luminosity and the characteristic time scales or post-peak decay rate of the light curves. Our model shows a correlation between higher peak luminosity and a faster decay of the light curve. This correlation between peak bolometric luminosity and $\Delta m_{15}$ is very tight; whereas the correlation between peak V-band magnitude and $\Delta m_{15}$ is less pronounced.
Importantly, the observed population of Type Ib/c supernovae does \emph{not} show such a correlation.
There may instead be a trend towards slower evolution of brighter events. This implies that the
strong viewing-angle dependence in our ultra-stripped model cannot be generic for stripped-envelope supernovae, otherwise these would have to show a much bigger scatter in $\Delta m_{15}$ for a given luminosity. Also, there is less of a tension in the scatter of $\Delta m_{15}$ versus peak V-band magnitude.

One possible (and interesting) explanation for this 
possible discrepancy would be that current 3D core-collapse supernovae are too asymmetric to be compatible with observed Ib/c supernova photometry. But it is also possible that the strong viewing-angle dependence is a peculiarity due to the low ejecta mass of
about $1\,\mathrm{M}_\odot$. Similar asymmetries of the $^{56}\mathrm{Ni}$ distribution may affect the light curve less if the $^{56}\mathrm{Ni}$ is hidden beneath a more massive envelope with higher optical depth. Given that the progenitor considered in this study still represents a rare stellar evolution channel with an unusually small envelope mass, further studies with more typical progenitors
are required to determine whether 3D core-collapse supernova explosion models are at odds with
the observed photometry of Ib/c supernovae.

Nonetheless, our 3D radiative transfer simulations show a promising avenue towards validating
or disproving 3D models of neutrino-driven supernovae of stripped-envelope progenitors. Asymmetries in the $^{56}\mathrm{Ni}$ distribution, which are an inherent consequence of the multi-dimensional dynamics seen in modern simulations of neutrino-driven explosions, can leave a significant viewing-angle dependence in the photometry of the observed transient. This viewing-angle dependence would have to be reflected in the distribution of photometric properties across the stripped-envelope supernovae population. The degree and direction of scatter of $\Delta m_{15}$ versus peak luminosity can then serve as a means to constrain explosion asymmetries, even without spectroscopic or polarimetric observations. Bigger controlled observational samples of stripped-envelope supernovae and 3D simulations for a wider range of progenitor models will be required to permit a quantitative comparison and constrain supernova explosion asymmetries.

\section*{Acknowledgements}
The authors acknowledge support by
the Australian Research Council (ARC)
through grants FT160100035 (BM), DP240101786 (BM, AH) and DP240103174.
AH also acknowledges support by
by the ARC Centre of Excellence for All Sky Astrophysics in 3 Dimensions (ASTRO 3D) through project number CE170100013, and by ARC LIEF grants LE200100012 and LE230100063.
FPC and SAS, acknowledge funding from STFC grant ST/X00094X/1.
SAS acknowledges support from the UK Science and Technology Facilities Council [grant numbers ST/P000312/1, ST/T000198/1, ST/X00094X/1].
We acknowledge computer time allocations from Astronomy Australia Limited's ASTAC scheme, the National Computational Merit Allocation Scheme (NCMAS), and
from an Australasian Leadership Computing Grant.
Some of this work was performed on the Gadi supercomputer with the assistance of resources and services from the National Computational Infrastructure (NCI), which is supported by the Australian Government, and through support by an Australasian Leadership Computing Grant.  Some of this work was performed on the OzSTAR national facility at Swinburne University of Technology.  OzSTAR is funded by Swinburne University of Technology and the National Collaborative Research Infrastructure Strategy (NCRIS). 

Some of this work used the DiRAC Memory Intensive service (Cosma8) at Durham University, managed by the Institute for Computational Cosmology on behalf of the STFC DiRAC HPC Facility (www.dirac.ac.uk). The DiRAC service at Durham was funded by BEIS, UKRI and STFC capital funding, Durham University and STFC operations grants. Some of this work also used the DiRAC Data Intensive service (CSD3) at the University of Cambridge, managed by the University of Cambridge University Information Services on behalf of the STFC DiRAC HPC Facility (www.dirac.ac.uk). The DiRAC component of CSD3 at Cambridge was funded by BEIS, UKRI and STFC capital funding and STFC operations grants. DiRAC is part of the UKRI Digital Research Infrastructure. The authors gratefully acknowledge the Gauss Centre for Supercomputing e.V. (www.gauss-centre.eu) for funding this project by providing computing time on the GCS Supercomputer JUWELS at Jülich Supercomputing Centre (JSC). NumPy and SciPy \citep{oliphant2007a}, Matplotlib \citep{hunter2007a}  and \href{https://zenodo.org/records/8302355} {\textsc{artistools}}\footnote{\href{https://github.com/artis-mcrt/artistools/}{https://github.com/artis-mcrt/artistools/}} \citep{artistools2024a} were used for data processing and plotting.

\section*{Data Availability}
The data from our simulations will be made available upon reasonable requests made to the authors.

\bibliographystyle{mnras}
\bibliography{paper}

\bsp
\label{lastpage}
\end{document}